\documentclass[aps,prb,twocolumn,amsmath,amssymb,longbibliography]{revtex4-2}

\usepackage{graphicx} 
\usepackage{setspace}
\usepackage{tikz}
\usepackage{amsmath}
\usetikzlibrary{positioning}
\usetikzlibrary{calc, positioning}
\usepackage{braket}
\usepackage{bm}
\usepackage{epstopdf}
\usepackage{multirow}
\usepackage{array}
\usepackage{relsize}

\begin{document}


\title{Quantum Localization in Incommensurate Tight-Binding Chains}

\author{C. J. Dyrseth}
\email{cd20md@brocku.ca}
\author{K. V. Samokhin}
\affiliation{Department of Physics, Brock University, St. Catharines, Ontario L2S 3A1, Canada}

\date{\today}

\begin{abstract}
We explore quantum localization phenomena in a system of two coupled tight-binding chains with incommensurate periods. Employing the inverse participation ratio as a measure of localization, we investigate the effects of geometric incommensurability and external magnetic fields. Numerical results reveal the existence of a mobility edge in the spectrum characterized by an abrupt onset of localization in higher-energy states. We find that localization tends to be enhanced by a weak magnetic field, whereas a strong field delocalizes most states.
\end{abstract}

\maketitle


\section{Introduction}
\label{sec: intro}

Electronic properties of disordered and quasi-periodic materials have long been the focus of extensive research in condensed matter physics. In periodic crystals, where atoms are arranged in regular, repeating patterns, the wave function can be expressed as the product of a plane wave and a function with the same periodicity as the lattice, a result known as Bloch's theorem \cite{Bloch1929berDQ}. Physically, this implies that the probability distribution of the electron is spread evenly throughout the lattice. However, in disordered or quasi-periodic materials, this result does not hold. In such systems, it is possible for the wave functions of electrons to become localized.

One of the first theoretical frameworks developed to study disordered materials is the Anderson model \cite{anderson1958absence, lee1985review,kramer1993review}. In this model, the transfer of electrons between nearest-neighbor sites of a crystal lattice is described using the tight-binding approximation \cite{ashcroft1976solid,kittel2004solid}, while the on-site energies are randomly selected from a specified range, with the extent of this range referred to as the disorder strength. In one dimension (1D), all electronic states become exponentially localized regardless of the disorder strength \cite{mott}. Various methods have been employed to analyze the Anderson model, including the transfer matrix methods \cite{transfer}, Green’s function techniques \cite{green}, and numerical simulations such as exact diagonalization and the Lanczos algorithm \cite{numerical}. These techniques calculate key physical quantities such as the localization length, density of states, and conductivity. Experimental validations of the Anderson model have been conducted using ultra-cold atom systems \cite{Billy_2008} and photonic lattices \cite{photon}. 

Localization in disordered systems usually tends to be suppressed by an external magnetic field. For instance, weakly localized systems in two dimensions (2D) exhibit negative magnetoresistance due to a reduction of constructive interference of the electronic wave functions caused by an additional magnetic phase shift \cite{Bergmann1984}. In the 2D Anderson model, a weak magnetic field reduces localization for states near the Landau band centers \cite{WeakField}.

A different class of systems, which are not randomly disordered but still lack perfect lattice periodicity, are quasi-periodic crystals. One of the most popular quasi-periodic 1D models is the Aubry-Andr\'e (AA) model \cite{articleAA,SOKOLOFF1985189,Dominguez_Castro_2019}, which features periodically varying on-site energies with the period incommensurate with the lattice, meaning the ratio between the two periods is irrational. In the AA model, all states become localized when the potential strength exceeds a certain threshold. The model is closely related to the Harper model, which describes electrons in a two-dimensional periodic lattice under a uniform magnetic field; the Harper equation can be mapped onto the AA model \cite{Harper:1955jqr}. The AA model can also be used to describe ultracold atoms in two incommensurate optical lattices. An experimental demonstration of the localization of bosonic atoms in such a system was reported in Ref. \cite{Roati_2008}, offering direct evidence of the AA model predictions. Further experiments have investigated many-body localization in quasi-periodic systems \cite{Schreiber_2015}.

Several models related to the AA model have been proposed to explore more complex localization phenomena in quasi-periodic systems. One such model is the Soukoulis–Economou model \cite{SouEco1982}, which introduces an additional potential term that breaks the self-duality of the original AA model. This modification allows the coexistence of both extended and localized states within the same system, separated by a mobility edge—critical energy value marking the transition between extended and localized behavior. Mobility edges have also been found theoretically in numerous other generalizations of the single-chain AA model \cite{GAA-1988,GAA-2009,GAA-2010,GAA-2015,GAA-2016,mosaic-2020,GAA-2023}. Experimental evidence of a mobility edge in a 1D quasiperiodic optical lattice has been reported in Ref. \cite{Lusch-2018}.

The Fibonacci chain \cite{Fib-chain-1986,Fib-chain-1987,Kohmoto1987,Jagan2021} provides yet another important example. In this 1D tight-binding model, the on-site potentials or hopping amplitudes follow a deterministic, quasi-periodic pattern based on the Fibonacci sequence. The resulting wavefunctions are critical, i.e. neither fully extended nor exponentially localized, instead decaying algebraically and exhibiting self-similar, fractal features. Interestingly, it has been shown that the AA model and the Fibonacci quasi-crystal are topologically equivalent, as one can be smoothly deformed into the other without closing any bulk energy gaps \cite{Kraus2012}.

In this paper, we construct a model of quasi-periodic crystals in which (i) incommensurability naturally arises from the geometry of the system and (ii) it is straightforward to study the effects of an external magnetic field on quasi-1D localization. Both goals can be achieved if one considers two coupled 1D chains of atoms, with the ratio $\rho$ of the lattice periods being an irrational number. We do not introduce any external on-site potentials, therefore, in contrast to the AA model and its multi-chain generalizations \cite{Guo-2014, moblity2,Lin-2023,two-leg-AA-2024, Kohlert,Li}, the incommensurability in our model affects only the hopping amplitudes. 

Ladder models or ``coupled moir\'e chains'' with geometry similar to that considered here have been studied previously \cite{WYS2021,similar2}. Those works focused on Gaussian distance-dependent inter-chain hopping with short-range intra-chain coupling and demonstrated mobility edges in such systems. In contrast, we treat hoppings between sites in both chains on equal footing and consider exponential distance-dependent hopping for both intra- and inter-chain, with geometrically enforced periodic boundary conditions. We further relate the mobility edge to gaps in the energy spectrum, and study the effects of an external magnetic field on localization.

For numerical analysis, we assume that the chains have different numbers of atoms but the same length, with periodic boundary conditions, and approximate $\rho$ by a ratio of two large coprime numbers.
To characterize localization of the electronic wave functions, we use the the inverse participation ratio (IPR) \cite{bell1970atomic,Thouless197493,wegner1980ipr}. The IPR takes values between zero and one, with the larger values corresponding to the more localized wave functions. 

The structure of the paper is as follows. In Sec.~II we introduce the model of two coupled incommensurate tight-binding chains. In Sec.~III we extend the model by including an applied magnetic field. In Sec.~IV we describe the numerical method and present the results. In Sec.~V we conclude with a discussion of the findings and their implications. A perturbative analysis of the effects of inter-chain hopping for weakly coupled chains is presented in the Appendix. Throughout the paper we use the units in which $c=1$ and $e$ is the absolute value of the electron charge.


\section{The Model}
\label{sec: model}

\newcommand{\figlab}[1]{\raisebox{1ex}{\textbf{(#1)}}}

\begin{figure}[t]
  \centering
  \figlab{a}\\[0.5ex]
  \includegraphics[width=0.4\textwidth]{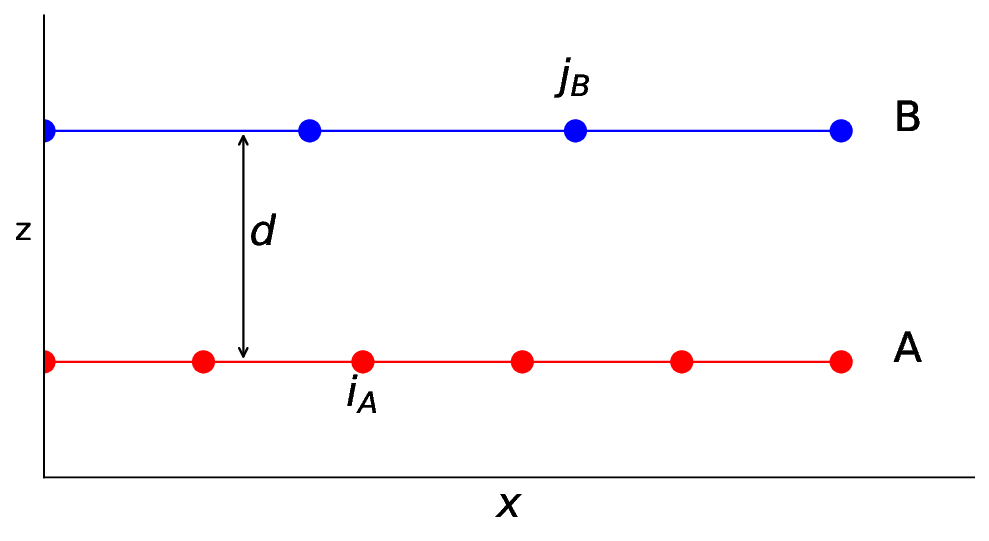}\\[1em]
  \figlab{b}\\[0.5ex]
  \includegraphics[width=0.4\textwidth]{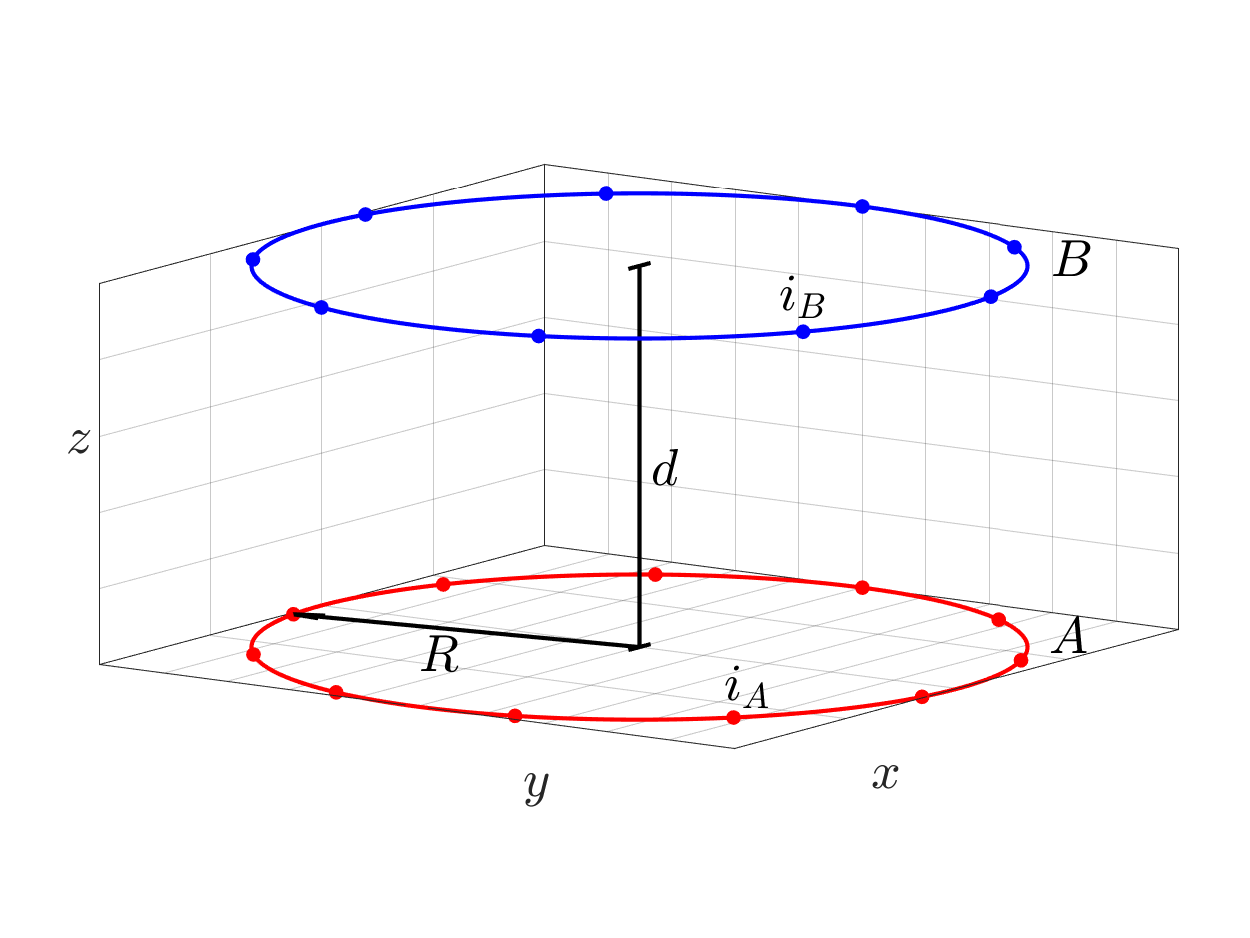}
  \caption{(Color online) Linear chains with periodic boundary conditions (a), represented by circular chains (b).  }
  \label{fig:coupled-chains}
\end{figure}

As a minimal model of a geometrically incommensurate 1D system, we consider two coupled chains, $A$ and $B$, of identical atoms. The chains have the same length, with the numbers of atoms in each chain denoted by $N_A$ and $N_B$. We assume periodic boundary conditions, so that our system essentially consists of two coupled tight-binding loops, as illustrated in Fig. \ref{fig:coupled-chains}. The circular geometry also allows one to naturally account for all possible hopping terms beyond nearest-neighbor ones. Although we are considering loops, for large enough systems the geometry will appear locally flat.  

The loops are vertically separated by a distance $d$ and share the same radius $R$. The coupling between the chains can be controlled by changing $d$. In particular, if $d$ is large enough we will have two independent chains. The atoms in each chain are equally spaced along the circumference, with arc lengths between adjacent atoms denoted by $a$ and $b$ for chains $A$ and $B$, respectively. These arc lengths satisfy the relation $N_A a = N_B b = 2\pi R$, so that one can also control the inter-atomic couplings by varying the number of atoms in each chain. 

To model an incommensurate system, we set the ratio 
\begin{equation}
\label{rho-def}
  \rho=\frac{N_A}{N_B}=\frac{b}{a} 
\end{equation}
to closely approximate an irrational number. This can be achieved using the method of continued fractions \cite{Azbel79,Sokoloff81}. We represent the ratio (\ref{rho-def}) in the following form:
\begin{equation}
    \rho = a_0 + \cfrac{1}{a_1 + \cfrac{1}{\ddots + \cfrac{1}{a_r}}},
\label{Eq:cont}
\end{equation}
where $a_0,a_1,\ldots,a_r$ are integers \cite{continue1964}. In our analysis we will try to bring $\rho$ close to the golden ratio $\phi=(1+\sqrt{5})/2\simeq 1.618$, which corresponds to all $a_i=1$ ($i\geq 0$). Truncating this infinite sequence to $r$ terms leads to a rational approximation for $\phi$ called the $r$th convergent and denoted as $\phi_r$. 
The latter is given by a ratio of consecutive Fibonacci numbers:
\begin{equation}
\label{phi-p-Fibo}
    \phi_r=\frac{F_{r+1}}{F_r}.
\end{equation}
The reciprocal of the golden ratio $\phi^{-1}=(\sqrt{5}-1)/2$, which is obtained by setting \( a_0 = 0 \) and \( a_i = 1 \) for \( i > 0 \), is also often used in the literature on incommensurate models. Using $\phi^{-1}$ effectively swaps the $A$ and $B$ chains, i.e., $N_A\leftrightarrow N_B$ and $a\leftrightarrow b$.

To develop a tight-binding model of the system, we introduce the position basis defined by the states \( |A, i_A \rangle \) and \( |B, i_B \rangle \), where $i_A=1,...,N_A,\quad i_B=1,...,N_B$ 
are the indices that label atoms in the chains $A$ and $B$, respectively. The Hamiltonian consists of three components: intra-chain hopping terms \( \hat{H}_A \) and \( \hat{H}_B \), and inter-chain hopping term \( \hat{\tilde H} \):
\begin{equation}
\label{eq:H-def}
    \hat H=\hat H_A+\hat H_B+\hat{\tilde H}.
\end{equation}
The intra-chain blocks are given by
\begin{equation} 
\label{eq:HAHB}
    \begin{array}{l}
        \hat{H}_A=-\sum\limits_{i_A\neq j_A} t_A(i_A,j_A) \ket{A,i_A}\bra{A,j_A}, \medskip \\
        \hat{H}_B=-\sum\limits_{i_B\neq j_B} t_B(i_B,j_B) \ket{B,i_B}\bra{B,j_B},
    \end{array} 
\end{equation}
where $t_A$ and $t_B$ are the hopping amplitudes within the $A$ and $B$ chains, respectively. The inter-chain blocks are given by 
\begin{align} 
\label{eq:HAB}
    \hat{\tilde H}&=-\sum_{i_A,j_B} \tilde t(i_A,j_B) \ket{A,i_A}\bra{B,j_B}+\mathrm{H.c.},   
\end{align}
where $\tilde t$ are the hopping amplitudes between the chains. 

In our model, we include the hopping amplitudes between \textit{all} pairs of atoms. Assuming, for simplicity, that there is one isotropic ($s$-wave) orbital per atom, the electron wave functions' overlap for two atoms located at $\bm{r}$ and $\bm{r}'$ depends only on the distance $|\bm{r}-\bm{r}'|$. Then, one can write 
\begin{eqnarray}
\label{eq:t-AA-BB-AB}
    && t_A(i_A,j_A) = t  \exp{\left(-\frac{|\bm{r}_A(i_A)-\bm{r}_A(j_A)|}{\lambda}\right)}, \nonumber \\
    && t_B(i_B,j_B) = t  \exp{\left(-\frac{|\bm{r}_B(i_B)-\bm{r}_B(j_B)|}{\lambda}\right)}, \nonumber \\ 
    && \tilde t(i_A,j_B) = t  \exp{\left(-\frac{|\bm{r}_A(i_A)-\bm{r}_B(j_B)|}{\lambda}\right)}, 
\end{eqnarray}
where $t$ is the hopping constant, which has the dimension of energy, and $\lambda$ characterizes the spatial extent of the atomic wave functions.      
The positions of atoms in the chains are given by
\begin{eqnarray*}
    && \bm{r}_A(i_A)=R\cos\frac{2\pi i_A}{N_A}\,\hat{\bm x}+R\sin\frac{2\pi i_A}{N_A}\,\hat{\bm y},\\
    && \bm{r}_B(i_B)=R\cos\frac{2\pi i_B}{N_B}\,\hat{\bm x}+R\sin\frac{2\pi i_B}{N_B}\,\hat{\bm y}+d\,\hat{\bm z},
    \label{Eq:position}
\end{eqnarray*}
see Fig. \ref{fig:coupled-chains}(b). Therefore, the hopping amplitudes take the following form:
\begin{eqnarray}
    && t_A(i_A,j_A)=t\exp\left[-\frac{2R}{\lambda}\left|\sin\frac{\pi(i_A-j_A)}{N_A}\right|\right],\nonumber\\
    && t_B(i_B,j_B)=t\exp\left[-\frac{2R}{\lambda}\left|\sin\frac{\pi(i_B-j_B)}{N_B}\right|\right],\nonumber \\
    && \tilde t(i_A,j_B) \nonumber\\
    && =t\exp\left[-\frac{2R}{\lambda}\sqrt{\sin^2\left(\frac{\pi i_A}{N_A}-\frac{\pi j_B}{N_B}\right)+\left(\frac{d}{2R}\right)^2}\right].\qquad
\label{eq:t-AA-BB-AB-1} 
\end{eqnarray}
Note that $t_A$ and $t_B$ are invariant with respect to the lattice translations in the respective chains, i.e., they depend only on $|i_A-j_A|$ or $|i_B-j_B|$, but have different magnitudes if $N_A\neq N_B$. In contrast, the inter-chain hopping amplitudes are \textit{not} translationally invariant.  

Although all possible hopping terms are included in our model, in practice, for any given atom, only a few neighbors in both chains make a non-negligible contribution. The magnitudes of the inter-chain hoppings vary from atom to atom in a quasi-random fashion, producing a ``geometrical disorder'' in the system.
We would like to mention here the study of the energy spectrum in a model of two coupled weakly incommensurate ``Moir\'e chains'' \cite{Moire-chain-2020}. In this model, only the nearest-neighbor intra-chain hopping amplitudes (same in both chains) were considered and a different dependence of the inter-chain hoppings on the distance between the sites was assumed. 


\subsection{Effects of magnetic field}
\label{sec: model with B}

We now include the effects of an external magnetic field in our tight-binding model, which can be achieved by using the Peierls substitution \cite{PhysRevB.14.2239, marder2010condensed}. In this method each hopping amplitude in Eqs. (\ref{eq:HAHB}) and (\ref{eq:HAB}) is modified by a complex phase factor: $t_{\bm{r}\bm{r}^\prime} \xrightarrow{} t_{\bm{r}\bm{r}^\prime} e^{i\varphi_{\bm{r}\bm{r}^\prime}}$, where
\begin{equation}
\label{eq:magnetic-phase}
    \varphi_{\bm{r}\bm{r}^\prime} = -\frac{e}{\hbar} \int_{\bm{r}}^{\bm{r}^\prime} \bm{A} \cdot d\bm{l}.
\end{equation}
The integration here is performed along the straight path the electron hops.

\begin{figure}[t]
  \centering
  \includegraphics[width=0.45\textwidth]{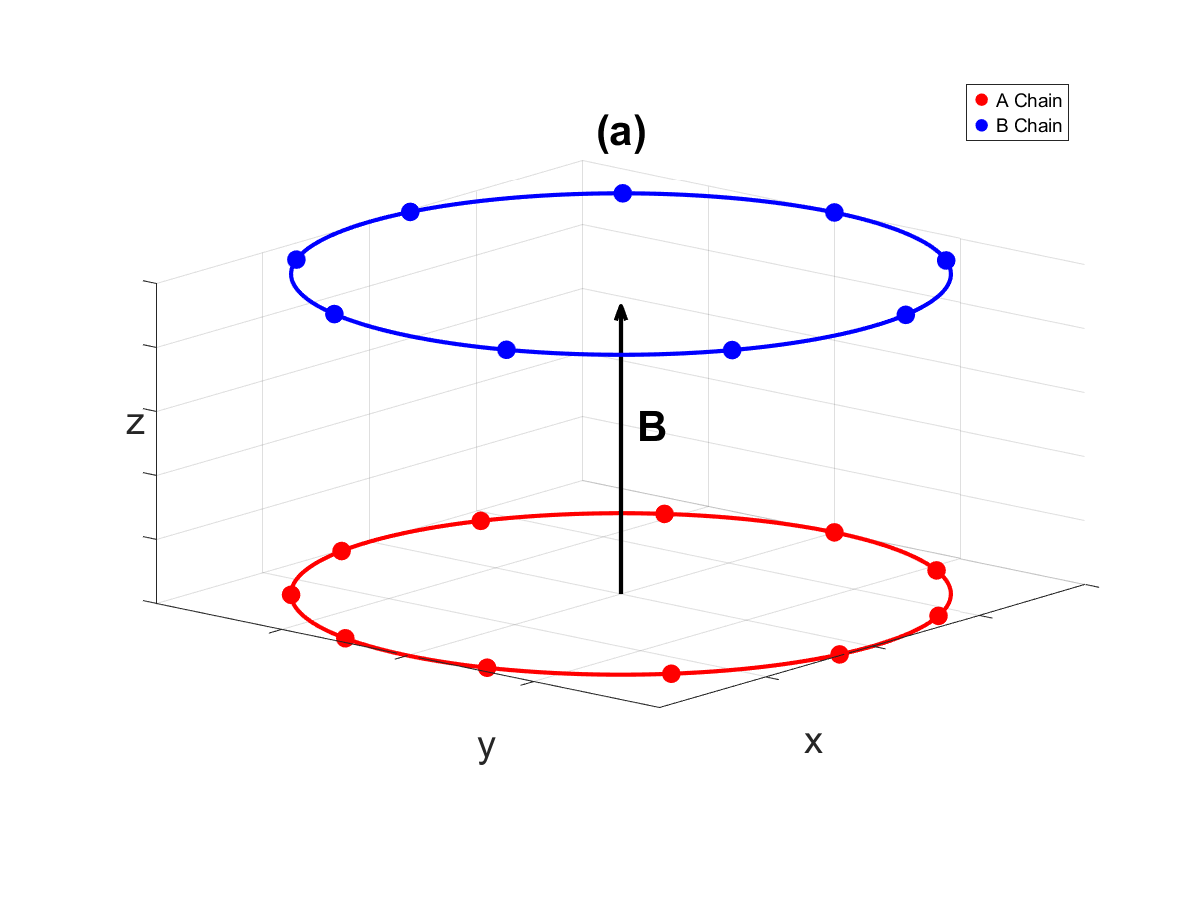}\\[1em]
  \includegraphics[width=0.30\textwidth]{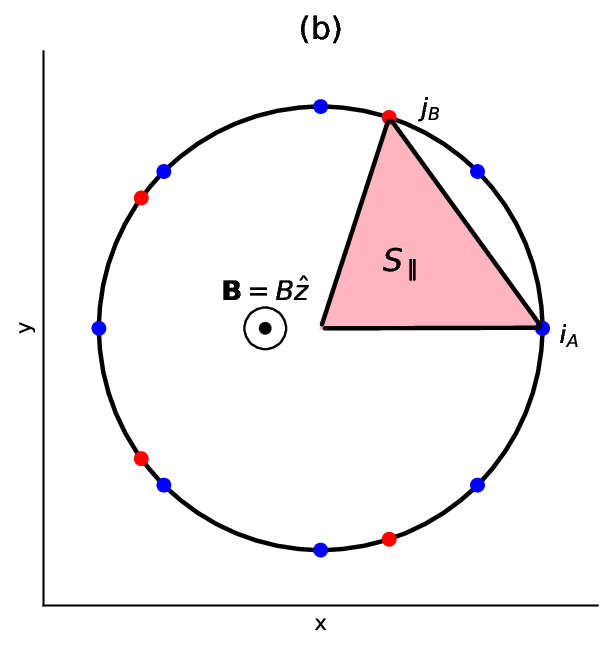}
  \caption{(Color online) (a) Circular chains with an external magnetic field orientated along the $z$-axis. (b) The triangular segment we integrate over in Eq. (\ref{Eq:radA}). The sites can be in different chains, as shown, or in the same chain.}
   \label{fig:ABcirc}
\end{figure}

Let us first consider an external magnetic field parallel to the chains' plane, i.e. oriented along the \( z \)-axis in the circular geometry, \( \bm{B} = B_\parallel \hat{\bm{z}} \), see Fig.~\ref{fig:ABcirc}. Using the cylindrical gauge for the vector potential, we have $\bm{A} = (B_\parallel R/2)\hat{\bm{\theta}}$.
Applying Stokes' theorem, it is straightforward to demonstrate that the line integral in Eq. (\ref{eq:magnetic-phase}) between sites \( i \) and \( j \) in the same or different chains is equal to the magnetic flux through the triangular area shown in Fig.~\ref{fig:ABcirc}. Namely,
\begin{equation}
    \varphi_{ij} = -\frac{1}{\Phi_0} \int_{S_\parallel} \bm{B} \cdot d\bm{S},
    \label{Eq:radA}
\end{equation}
where \( \Phi_0 = \hbar / e \) is the magnetic flux quantum. The explicit forms of the phases for intra-chain and inter-chain hopping are given by
$$
    \varphi(i_A,j_A) = \chi_\parallel \frac{N_A}{2\pi} \sin\frac{2\pi(i_A-j_A)}{N_A},
$$
$$
    \varphi(i_B,j_B) = \chi_\parallel \frac{N_A}{2\pi} \sin\frac{2\pi(i_B-j_B)}{N_B}, 
$$
$$
    \varphi(i_A,j_B) = \chi_\parallel \frac{N_A}{2\pi} \sin{\left(\frac{2\pi}{N_A}i_A-\frac{2\pi}{N_B}j_B\right)}.
$$
Here
\begin{equation}
    \chi_\parallel = \frac{\pi R^2B_\parallel}{N_A\Phi_0},
    \label{Eq:chiz}
\end{equation}
is the dimensionless magnetic field strength, which will serve as an adjustable parameter in our numerical analysis.

\begin{figure}[t]
  \centering
  \includegraphics[width=0.45\textwidth]{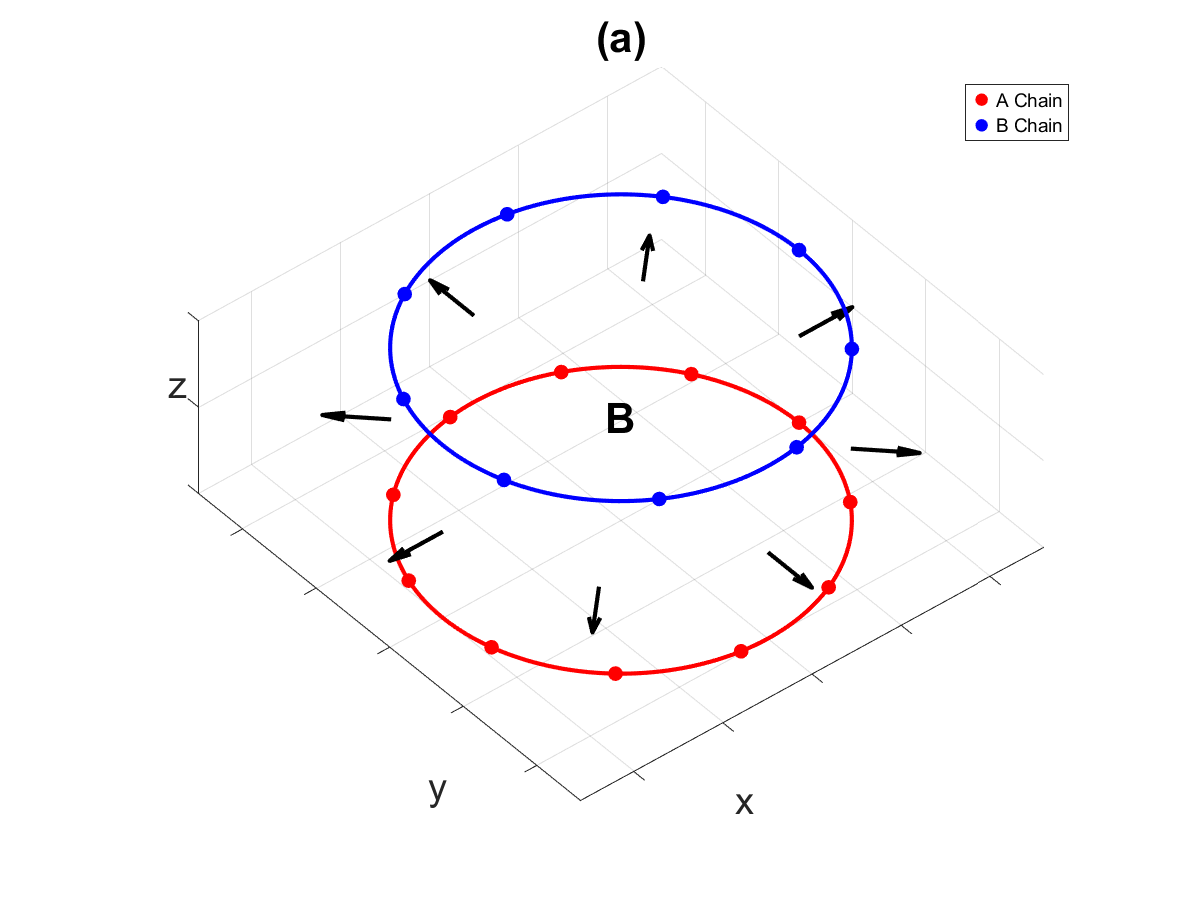}\\[1em]
  \includegraphics[width=0.32\textwidth]{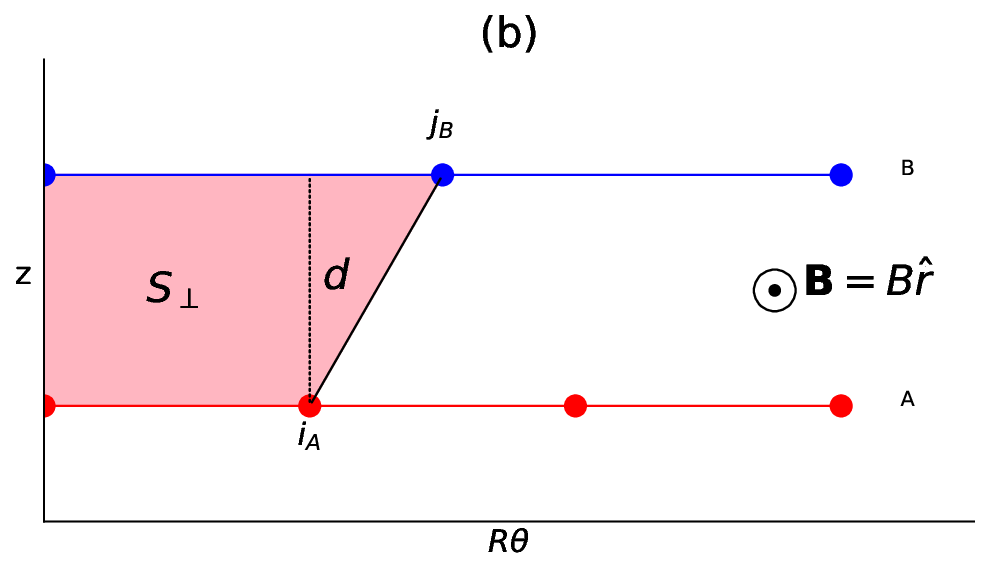}
  \caption{(Color online) (a) Circular chains with an external magnetic field orientated along the normal of the surface of the cylinder formed by the two chains. (b) The incommensurate chain system in a perpendicular (radial) magnetic field. The shaded area shows the segment we integrate over in Eq. (\ref{Eq:area_int}).}
  \label{fig:ABflux}
\end{figure}

Next, we investigate the effects of a radial magnetic field \( \bm{B} = B_\perp \hat{\bm{r}} \) on the two tight-binding loops. Note that a purely radial magnetic field would violate Gauss's law for magnetism $\bm{\nabla} \cdot \bm{B}=0$ and therefore cannot exist in nature. We use this field configuration to approximate two long chains with periodic boundary conditions under a \textit{perpendicular} field as depicted in Fig. \ref{fig:ABflux}.

The vector potential corresponding to the radial field can be chosen as \( \bm{A} = B_\perp R\theta \hat{\bm{z}} \), where $\theta$ is the polar angle in the $xy$ plane. It is easy to see that only the inter-chain hopping amplitudes are affected by a phase factor, which is given by
\begin{equation}
    \varphi(i_A,j_B) = -\frac{1}{\Phi_0} \int_{S_\perp} \bm{B} \cdot d\bm{S}.
\label{Eq:area_int}
\end{equation}
This integral represents the magnetic flux through the shaded area $S_\perp$ shown in Fig. \ref{fig:ABflux}. The resulting phase factor becomes
\begin{equation}
    \varphi(i_A,j_B) = -\chi_\perp (i_A + \rho j_B), \nonumber   
\end{equation}
where
\begin{equation}
    \chi_\perp = \frac{\pi B_\perp Rd}{N_A\Phi_0},  
\label{Eq:chir}   
\end{equation}
is the dimensionless strength of a perpendicular magnetic field.


\section{Numerical Analysis}
\label{sec: numerics}

In periodic systems, Bloch's theorem enables analytical calculation of the energy spectrum by assuming extended plane wave solutions for the wave function. However, our system does not repeat periodically; instead, it consists of $N_A + N_B$ atoms within a single ``unit cell'', which coincides with the whole crystal. This non-periodicity means that extended plane-wave solutions are no longer valid and that Bloch's theorem cannot be applied, resulting in $N_A + N_B$ distinct energy levels rather than energy bands. By increasing the size of the system and tuning the ratio $\rho$ to approximate an irrational number, we effectively render the system incommensurate. Since there is no analytical solution available for such a system, we must resort to numerical methods to find the wave functions and energy eigenvalues.

We cannot label the energy levels in terms of the wave vector $\bm{k}$ as in a periodic system. Instead, we now label the energy levels from lowest to highest using an integer index $n=1,...,N_A+N_B$. The eigenstates of the Hamiltonian are denoted as $\ket{n}$, and the corresponding energies -- as $E_n$. Measuring the latter in units of the hopping constant $t$, see Eq. (\ref{eq:t-AA-BB-AB}), we solve the following eigenvalue problem:
\begin{equation}
    \frac{\hat{H}}{t} \ket{n} = E_n \ket{n}, 
\label{Eq:Eigen}
\end{equation}
where $\hat{H}$ is given by Eq. (\ref{eq:H-def}). The eigenvectors have the following form:
\begin{equation}
    \ket{n}=\begin{pmatrix}
\psi_n(A,1) \\
\vdots \\
\psi_n(A,N_A)\\
\psi_n(B,1)\\
\vdots \\
\psi_n(B,N_B)
\end{pmatrix},
\label{Eq:eigenstate}
\end{equation}
with the components $\psi_n(A,i_A)=\braket{A,i_A|n}$ and $\psi_n(B,i_B)=\braket{B,i_B|n}$. The eigenvectors are normalized:
$$
    \braket{n|n}=\sum_{i_A=1}^{N_A}|\psi_n(A,i_A)|^2+\sum_{i_B=1}^{N_B}|\psi_n(B,i_B)|^2=1.
$$

In our model, following a considerable precedent in the previous studies of incommensurate 1D systems \cite{articleAA}, we focus on using the golden ratio \( \phi \) as the incommensurability parameter. To better represent real physical systems we select large values for \( N_A \) and \( N_B \). By choosing values of \( N_A \) and \( N_B \) in the thousands, we ensure a high degree of precision in approximating $\phi$ while maintaining practical relevance for numerical analysis. Namely, we use
\begin{equation}
    \rho=\phi_{17}=\frac{2584}{1597},
\end{equation}
as determined from Eq. (\ref{phi-p-Fibo}), i.e. set $N_A=F_{18}=2584$ and $N_B=F_{17}=1597$, where $F_r$ is the $r$th Fibonacci number. Selecting higher-order approximations does not qualitatively affect our results.

To facilitate our analysis, we introduce dimensionless parameters that characterize the system's geometric properties. Measuring the lattice constant in the $A$ chain and the vertical spacing between the chains in units of the size of the atomic wave functions (or, in other words, setting the latter to unity for simplicity), we have
$a/\lambda\to a$ and $d/\lambda\to d$. From Eq. (\ref{eq:t-AA-BB-AB-1}), we obtain the dimensionless hopping amplitudes
\begin{eqnarray}
\label{eq:t-dimensionless}
    && t_A(i_A,j_A)=\exp{\left[-\frac{N_Aa}{\pi}\left|\sin\frac{\pi(i_A-j_A)}{N_A}\right| \right]}, \nonumber \\
    && t_B(i_B,j_B)=\exp{\left[-\frac{N_Bb}{\pi}\left|\sin\frac{\pi(i_B-j_B)}{N_B}\right|\right]}, \nonumber \\ 
    && \tilde t(i_A,j_B) \nonumber \\
    && \quad=\exp{ \left[-\frac{N_Aa}{\pi}
    \sqrt{\sin^2\left(\frac{\pi i_A}{N_A}-\frac{\pi j_B}{N_B}\right)+\left(\frac{\pi d}{N_Aa}\right)^{2}}\right] }, \nonumber\\
    &&
\end{eqnarray}
where we used $N_Aa=N_Bb$.

As a measure of the wave function's extent over lattice sites we use the IPR, which yields the reciprocal of the count of sites encompassed by the wave function. The IPR of a normalized state $\ket{\psi}$ is defined as \cite{bell1970atomic,Thouless197493,wegner1980ipr}
$$
    \text{IPR}=\sum_{i=1}^N |\psi(i)|^4, 
$$
where $\psi(i)=\braket{i|\psi}$ is the value of the wave function at the $i^{\rm th}$ site. 
In our two-chain model, the above expression takes the form
\begin{equation}
\label{Eq:IPR}
    \text{IPR}(n)=\sum_{i_A=1}^{N_A}|\psi_n(A,i_A)|^4+\sum_{i_B=1}^{N_B}|\psi_n(B,i_B)|^4.
\end{equation}
The value of the IPR provides valuable insight into the degree of localization of a given quantum state. The IPR trends towards zero (scaling as $L^{-1}$, where $L$ is the chains' length) for a delocalized wave function, whereas it is equal to unity in the case of a wave function confined to a single lattice site. 


\section{Results}
\label{sec: results}

In our numerical calculations, we use $N_A=2584$ and $N_B=1597$ and diagonalize a $4181\times 4181$ Hamiltonian matrix for different values of the dimensionless lattice constant $a$ and the inter-chain separation $d$ (the lattice constant in the $B$ chain is given by $b=\rho a$). The effects of a magnetic field, which are described by the dimensionless parameters $\chi_\parallel$ and $\chi_\perp$, are considered in Sec. \ref{sec: results with B}.

\subsection{Energy spectrum}

\begin{figure}[t]
\includegraphics[width=0.4\textwidth]{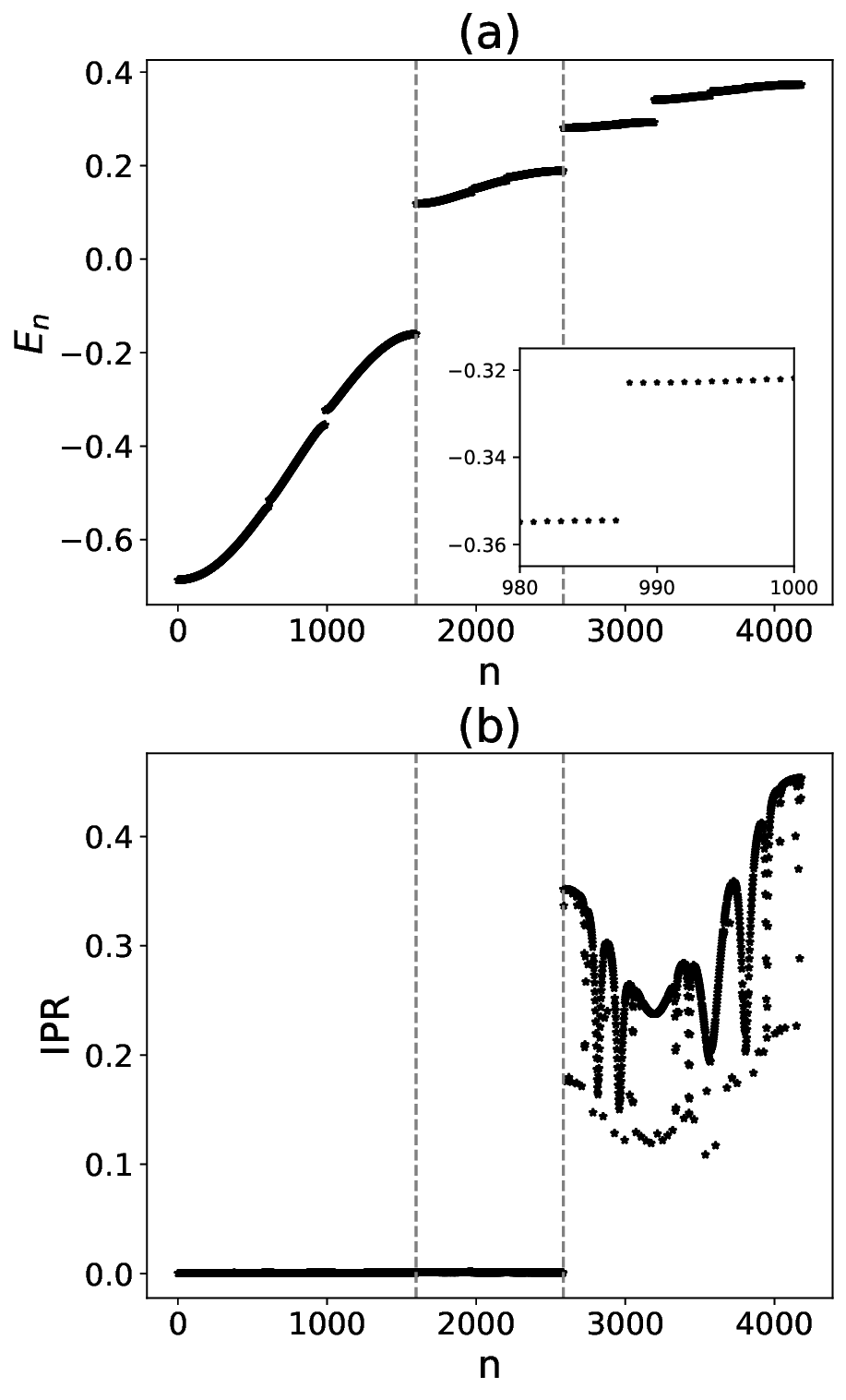}
  \caption{Data collected from a system with \(a=2\) and \(d=1\). (a) The energy spectrum, ordered from lowest to highest, is plotted over the energy index \(n\) ($1\leq n\leq N_A+N_B$). (b) The IPR plotted over the energy index \(n\). The dashed lines show the locations of large energy gaps at \(n=N_B\) and \(n=N_A\).}
  \label{fig:4gstop}
\end{figure}

In Fig.~\ref{fig:4gstop}, we present the energy spectrum and corresponding values of the IPR for a set of parameters where localization is observed, including a ``zoomed in'' view of one segment from the energy spectrum. The spectrum comprises discrete energy levels, and notably, it exhibits several gaps that are significantly wider than the typical spacing between adjacent energy levels.  It is possible to average our results with respect to the rotation of one of the chains. For instance, we can average with respect to the rotation of the $B$ chain through an angle of $0$ to $2\pi/N_A$. However, since this has a negligible impact on our results, we have chosen not to average in this way.

\begin{figure}[t]
\includegraphics[width=0.48\textwidth]{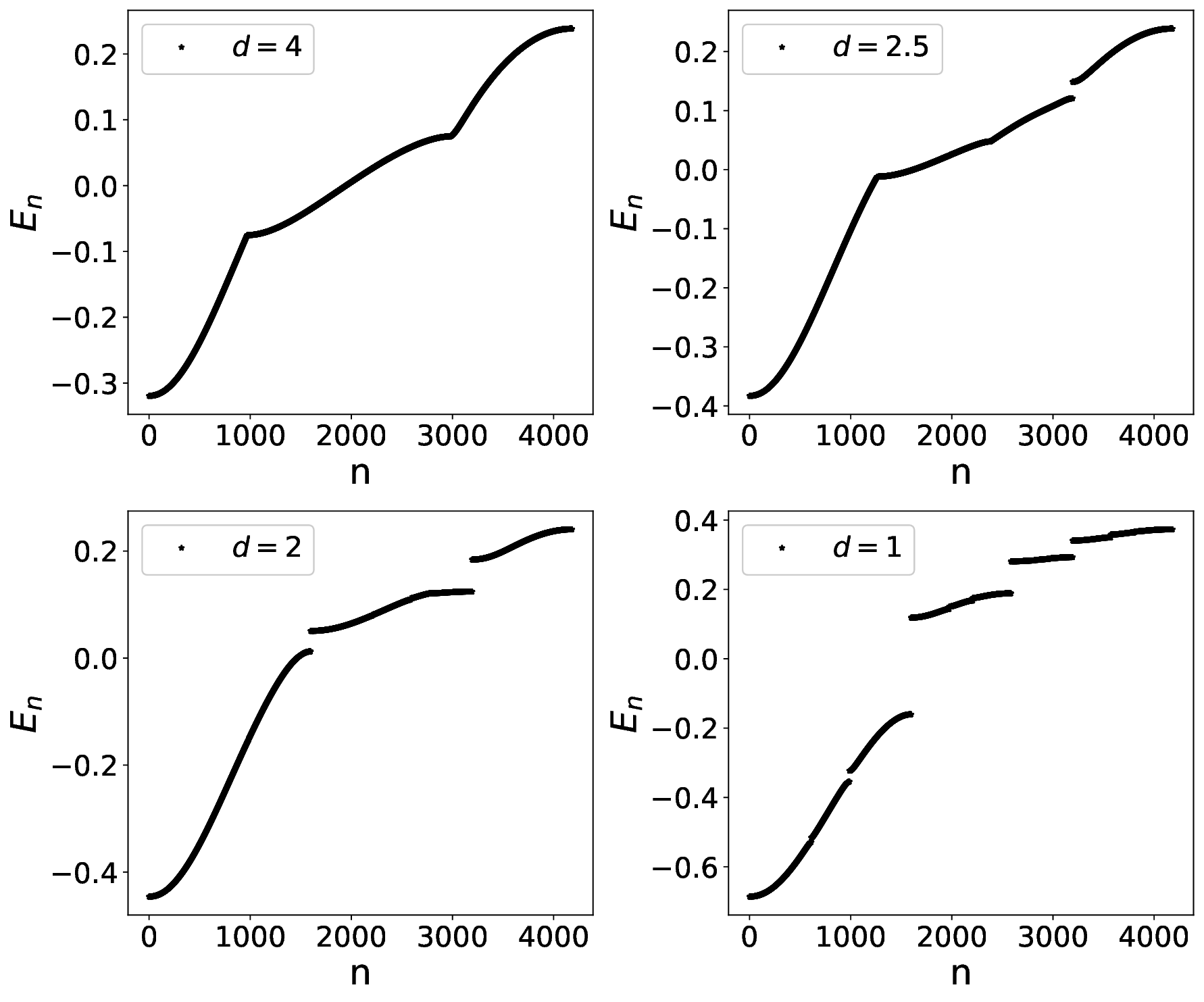}
  \caption{Evolution of the energy spectrum as the inter-chain separation $d$ is decreased at fixed $a=2$. }
  \label{fig:En_evolution}
\end{figure}

The evolution of the energy spectrum as a function of the inter-chain separation $d$ is shown in Fig.~\ref{fig:En_evolution}. At sufficiently large values of $d$ the spectrum is smooth and agrees with analytical results for decoupled chains. When $d$ is decreased, gaps begin to appear in the spectrum. Interestingly, these gaps seem to be located exactly at indices corresponding to the Fibonacci numbers $n=F_r$ or their simple combinations. This can be understood as an effect of ``small denominators'' in the perturbation expansions with respect to a weak inter-chain coupling, see Appendix \ref{app: decoupled chains}.

\begin{figure}[t]
  \centering
  \includegraphics[width=0.45\textwidth]{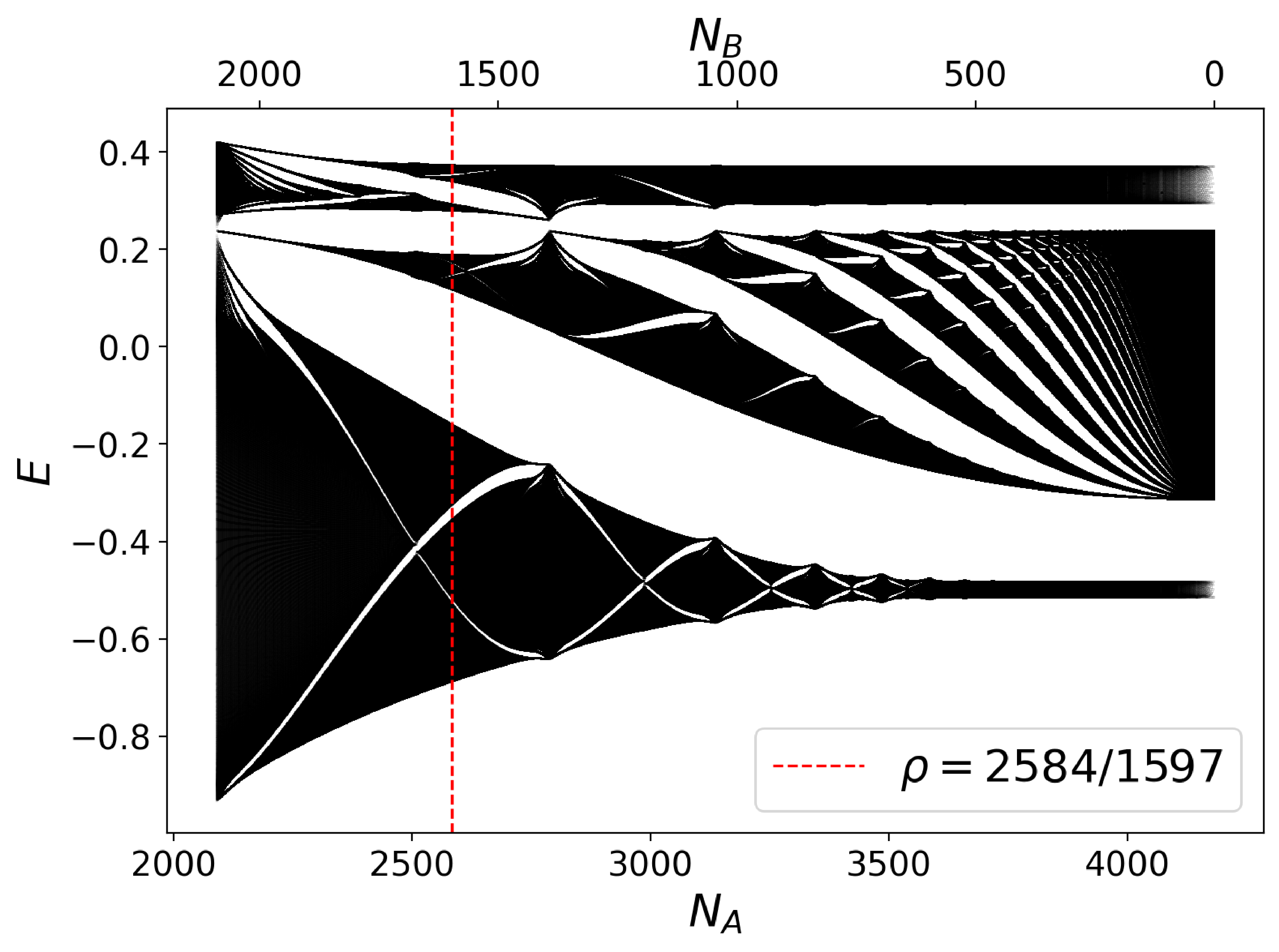}
  \caption{ (Color online) The energy spectrum for different ratios $\rho = N_A/N_B$, chosen according to Eq.~(\ref{Eq:N_A_rho}), for $N=4181$, $a=2$, and $d=1$. The vertical dashed line corresponds to $N_A=2584$ and $N_B=1597$. }
  \label{fig:E_rho_plots}
\end{figure}

Next, we examine how the incommensurability parameter $\rho$ defined in Eq.~(\ref{rho-def}) influences the energy spectrum. We will focus on the case $N_A \geq N_B$. To explore the full range of $\rho$, we fix the total number of sites in the system, $N = N_A + N_B$, and allow $N_A$ and $N_B$ to vary under this constraint. In this parametrization, $\rho$ can be written as
\begin{equation}
    \rho = \frac{N_A}{\,N - N_A\,},
    \label{Eq:N_A_rho}
\end{equation}
with $N/2 \leq N_A \leq N - 1$ (equivalently, $1 \leq N_B \leq N/2$), which restricts it to the interval $1 \leq \rho \leq N - 1$. Physically, varying $N_A$ in this way takes us from the ``maximally commensurate'' case where the two chains contain the same number of sites to the extreme limit where the $B$-chain contains only a single site. We plot the energy spectrum for each of these $(N_A,N_B)$ pairs in Fig.~\ref{fig:E_rho_plots}.

\subsection{IPR}

\begin{figure}[t]
\includegraphics[width=0.48\textwidth]{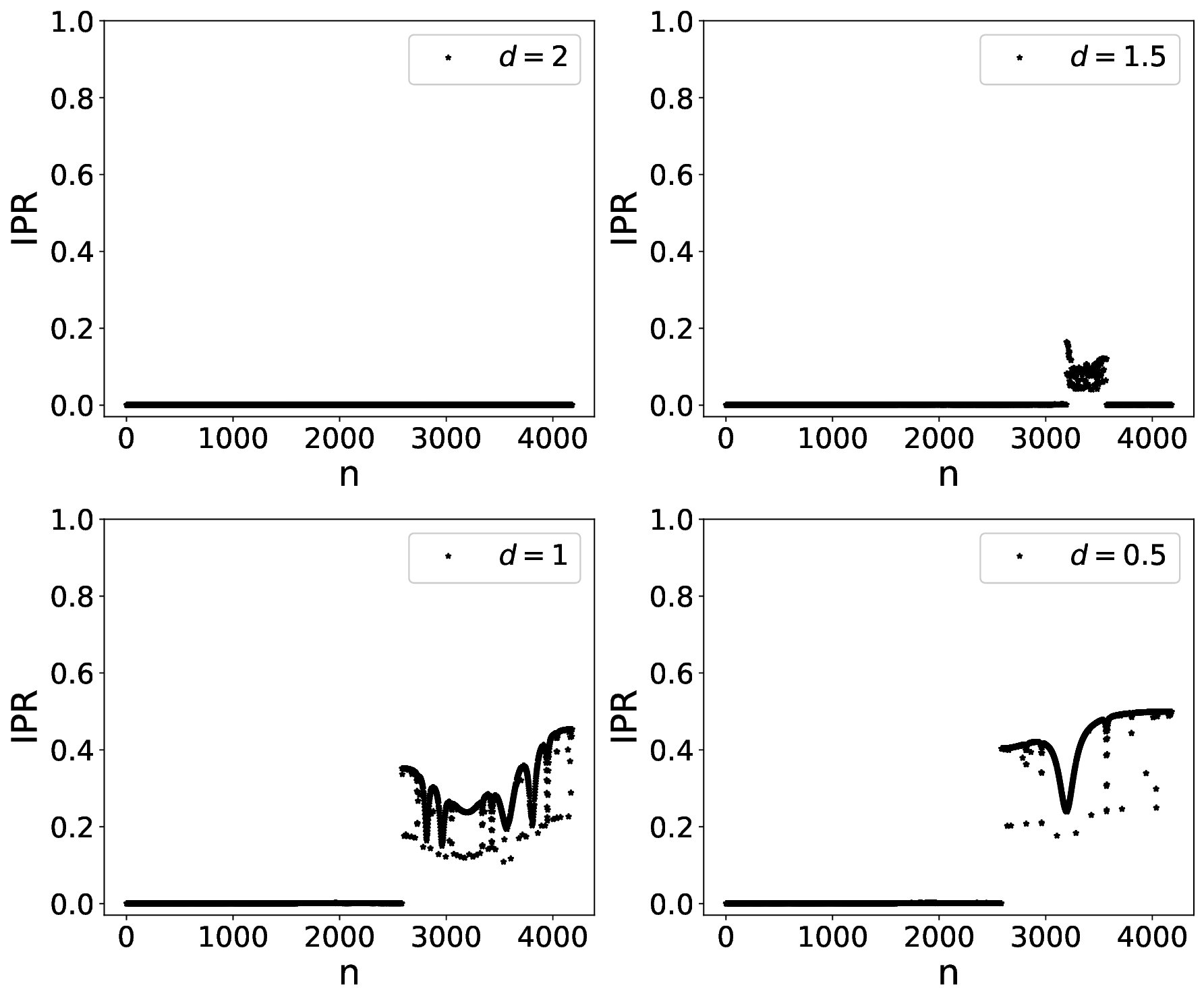}
  \caption{Evolution of the IPR as the inter-chain separation $d$ is decreased at fixed $a=2$. }
  \label{fig:tautilde}
\end{figure}

The relationship between the IPR and the inter-chain separation is shown in Fig.~\ref{fig:tautilde}. Qualitatively, there is an inverse relationship between the IPR and $d$, indicating that as $d$ increases, the IPR decreases. This confirms the intuitive expectation that decoupling the chains by increasing the vertical spacing between them reduces the amount of localization.

\begin{figure}[t]
\includegraphics[width=0.48\textwidth]{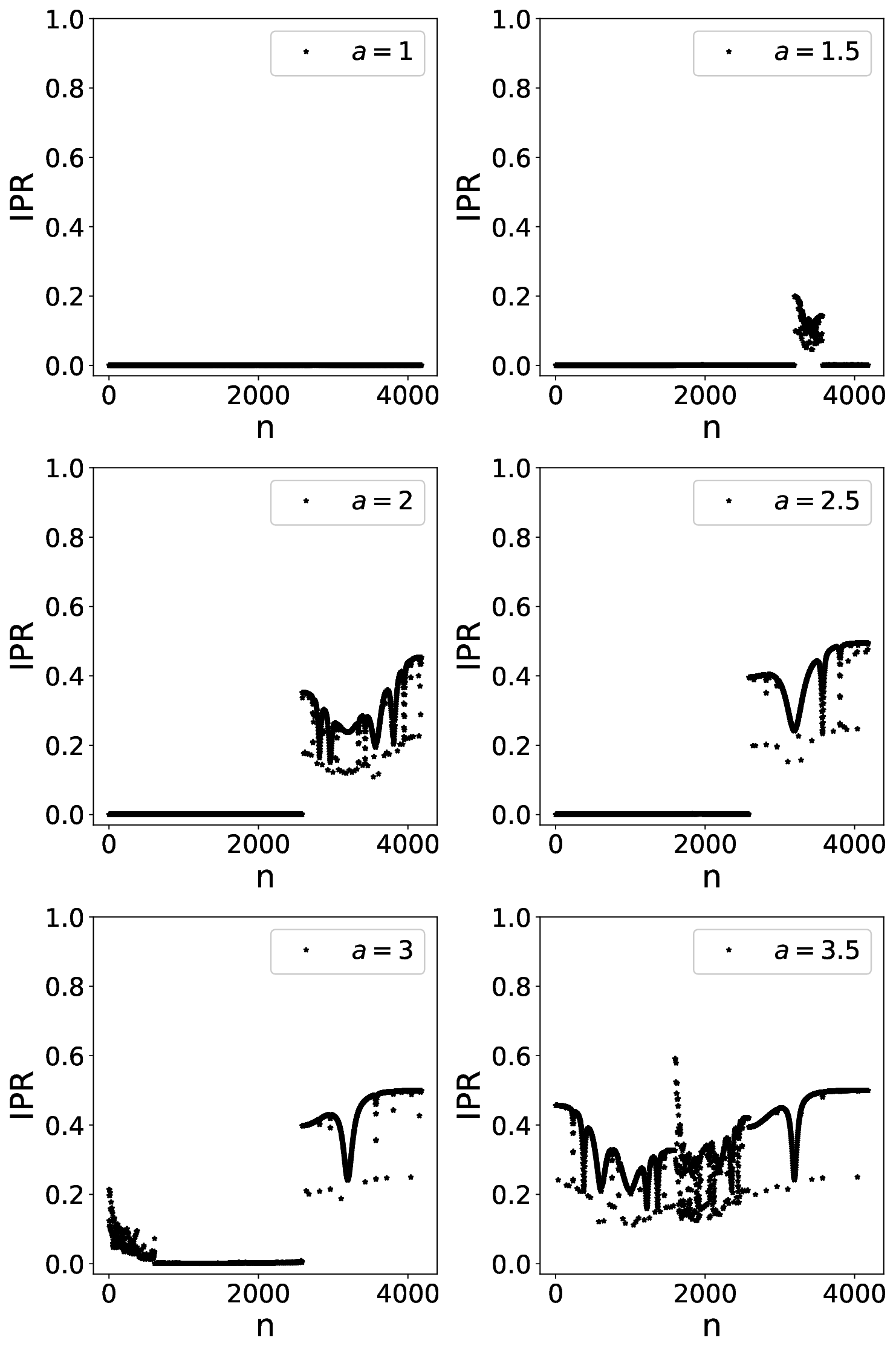}
  \caption{Dependence of the IPR on the lattice constant $a$ at fixed $d=1$.}
  \label{fig:tauA}
\end{figure}

The dependence of the IPR on the lattice constant $a$ is shown in Fig.~\ref{fig:tauA}. We see that increasing $a$ while keeping the inter-chain spacing constant decreases the prevalence of intra-chain hopping relative to inter-chain hopping, which in turn increases the number of localized states.

\begin{figure}[t]
\includegraphics[width=0.45\textwidth]{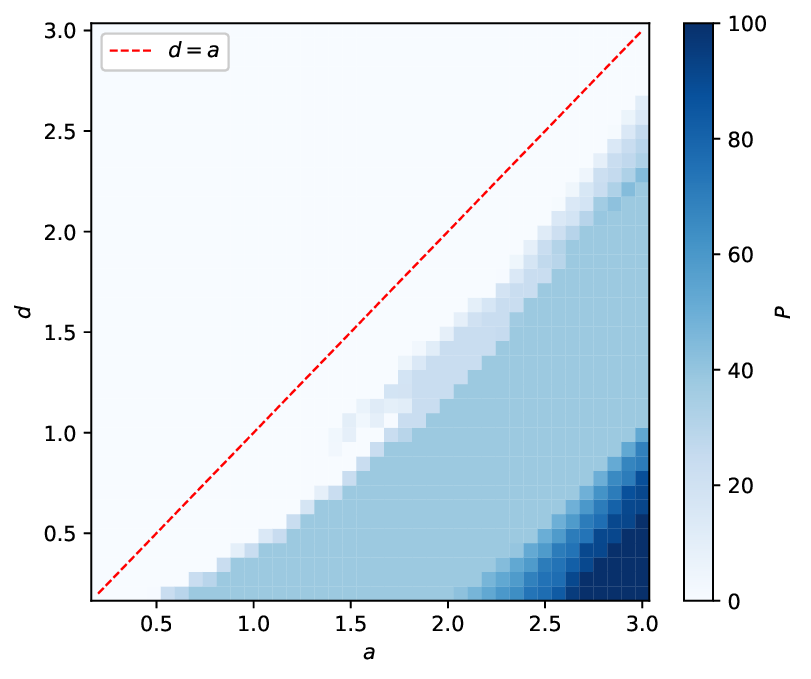}
  \caption{(Color online) The percentage of localized states, defined by the criterion $\text{IPR} > (N_A+N_B)^{-1/2} \approx 0.0154$, is shown for different values of $a$ and $d$. The dotted red line marks the boundary between regimes dominated by inter-chain and intra-chain hopping.}
  \label{fig:doubletau}
\end{figure}

The percentage $P$ of localized states in the system, for different values of the parameters $a$ and $d$, is shown in Fig.~\ref{fig:doubletau}. We define the threshold for localization as $\mathrm{IPR}>(N_A+N_B)^{-1/2}$, i.e. as the geometric mean between the IPRs for a fully extended state, $\mathrm{IPR}=(N_A+N_B)^{-1}$ and a fully localized state, $\mathrm{IPR}=1$. In the figure, the vertical axis represents $d$ and the horizontal axis represents $a$, while the red dashed line corresponds to $d = a$. Below the dotted line, inter-chain hopping dominates over intra-chain hopping, whereas above this line, intra-chain hopping is predominant. Notably, all localized states are found below the dotted line.

Our results for the coupled chains indicate that higher-energy states tend to be localized, exhibiting nonzero values of the IPR, whereas lower-energy states remain extended. Notably, the transition from extended to localized states is not continuous; instead, it appears to be abrupt and coincides with a gap in the energy spectrum. These findings suggest a strong correlation between the presence of energy gaps and the onset of localization in higher-energy states.

Plotting the IPR for individual eigenstates can be misleading, since the IPR of critical states also vanishes in the thermodynamic limit. To distinguish extended and localized states from critical ones, we therefore study how the IPR scales with the system size \cite{IPR_Scaling}. In general, the IPR scales as
\begin{equation}
\label{eq:IPR-scaling-N}
    \mathrm{IPR}(N) \sim N^{-D},
\end{equation}
where $N$ is the total number of sites in the system, in our case $N=N_A+N_B$. Extended states are characterized by $D = 1$, for which $\mathrm{IPR} \sim 1/N$, while localized states have $D = 0$, corresponding to an IPR that remains finite as $N$ increases. Critical states exhibit intermediate scaling with $0 < D < 1$; although their IPR also vanishes in the thermodynamic limit, these states are neither fully extended nor localized.

We focus on the parameter regime with $a = 2$, $d = 1$, and $\rho$ approximating the golden ratio. In this case, the spectrum contains a high-energy region with nonzero IPR and a low-energy region with vanishing IPR. The two regions are separated by a mobility edge, as shown in Fig.~\ref{fig:4gstop}(b). To characterize the nature of the states in each region, we compute the $\mathrm{IPR}$ averaged over a window of $0.01N$ eigenstates well inside each region, away from the mobility edge. We then analyze how the average $\mathrm{IPR}$ scales with the system size and extract the scaling exponent $D$, see Eq. (\ref{eq:IPR-scaling-N}). In accordance with Eq.~(\ref{phi-p-Fibo}), the system sizes are chosen such that $N_A$ and $N_B$ are consecutive Fibonacci numbers, ensuring that their ratio provides increasingly accurate rational approximations to the golden ratio.

\begin{figure}[t]
\includegraphics[width=0.35\textwidth]{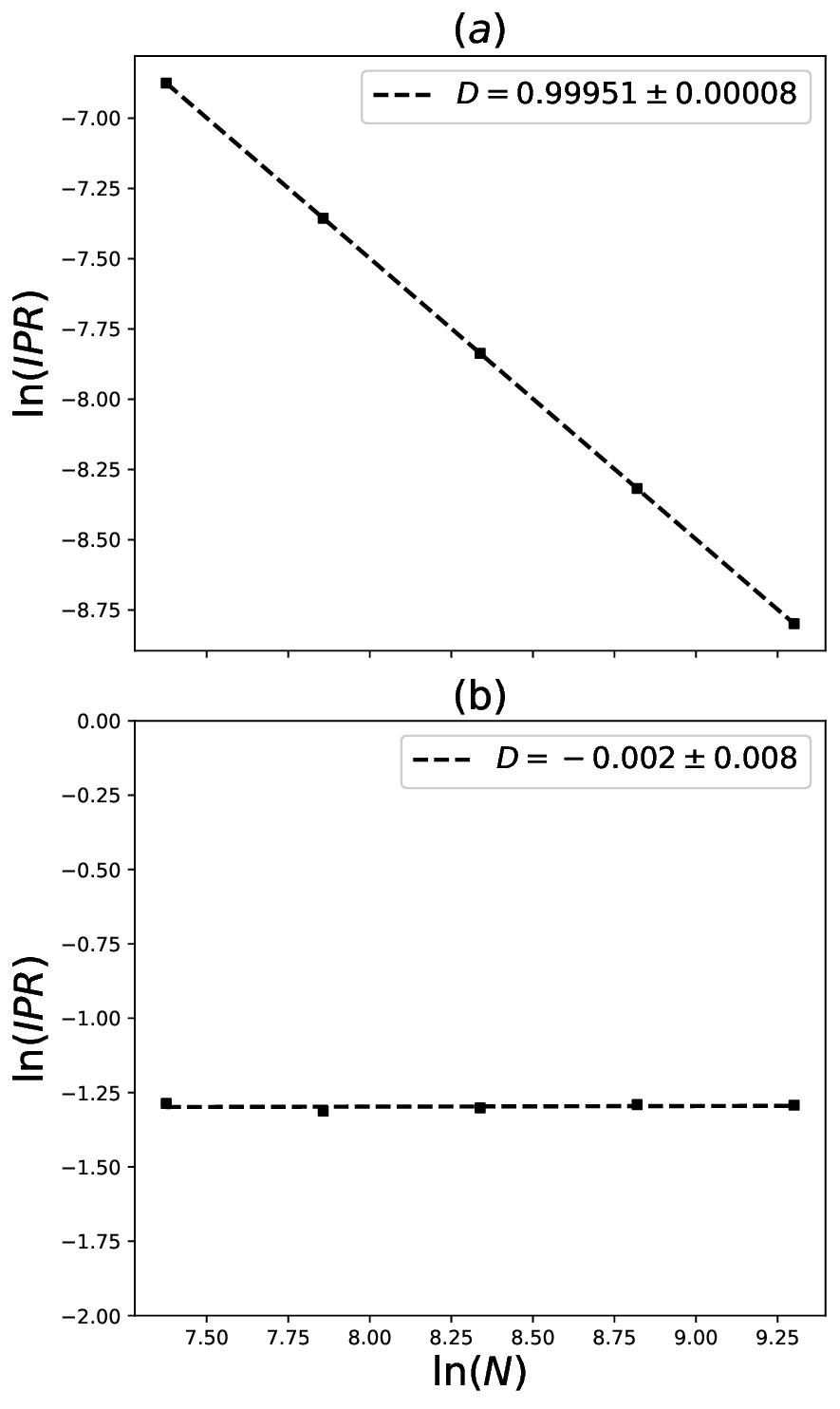}
  \caption{ Log-log plots of the average IPR vs the system size, fitted to a linear relation for $a=2$ and $d=1$. (a) Extended states in the low-energy region. (b) Localized states in the high-energy region. }
  \label{fig:IPR_Scaling}
\end{figure}

In Fig.~\ref{fig:IPR_Scaling}, we show the results of the IPR scaling analysis. In the low-energy region, we focus on the window of eigenstates centered at $n=N_A/2$ with width $0.01N$. From this analysis, we extracted a scaling exponent $D = 0.99951 \pm 0.00008$, which suggests that the states in this region are extended. In the high-energy region, we focus on the window of eigenstates centered at $n=N_A+N_B/2$ with width $0.01N$. In this case, we extracted the scaling exponent $D = -0.002 \pm 0.008$, which is equal to zero within statistical uncertainty, suggesting that the states in this region are indeed localized. In both energy regions we find that scaling the system does not affect the nature of the eigenstates.

\begin{figure}[t]
\includegraphics[width=0.35\textwidth]{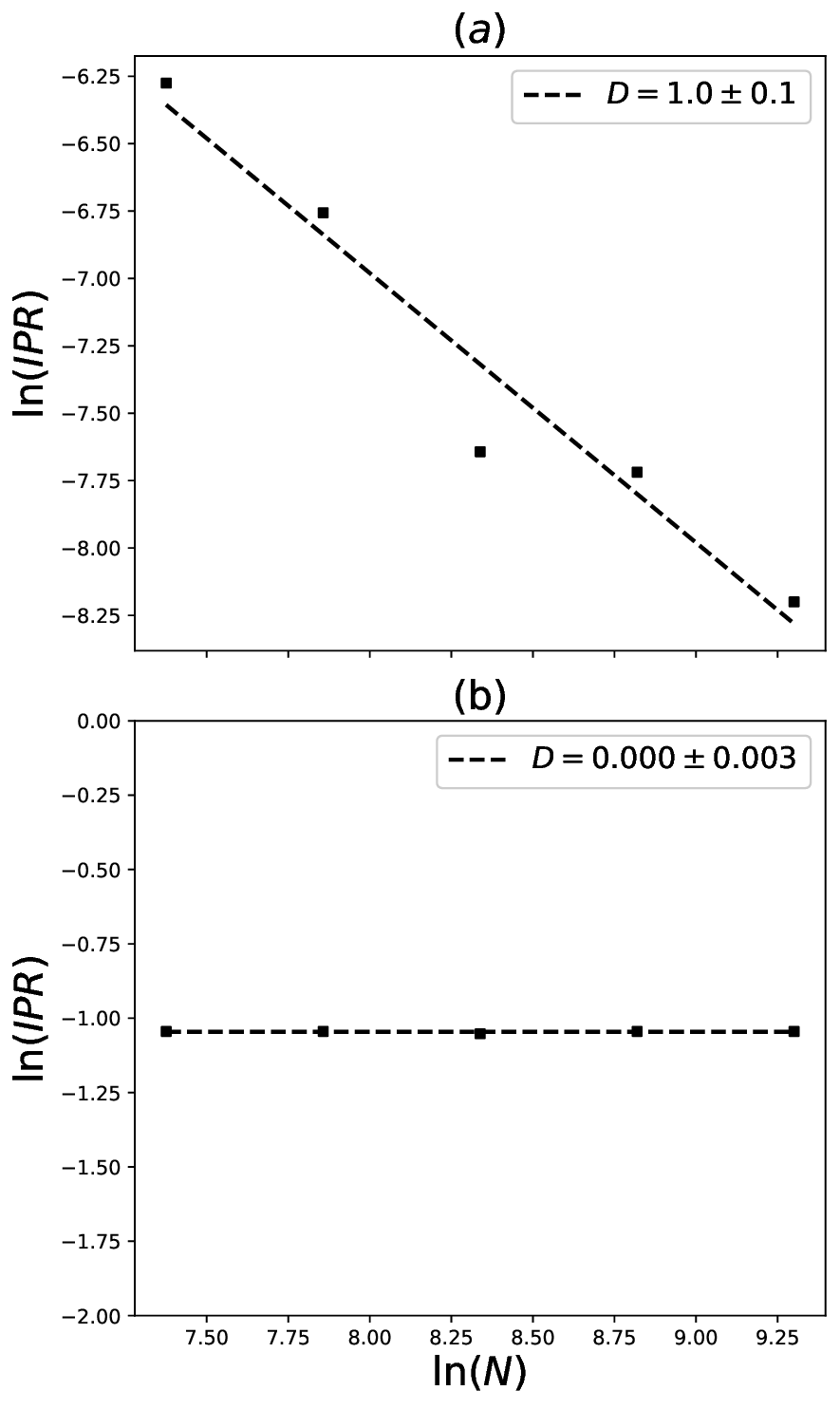}
  \caption{Log-log plots of the IPR vs the system size, fitted to a linear relation for $a=2$ and $d=1$. (a) Single state on the low-energy (extended) side of the mobility edge. (b) Single state on the high-energy (localized) side of the mobility edge. }
  \label{fig:IPR_Scaling_edge}
\end{figure}

Next, we characterize the states at the mobility edge. The state with $n = N_A$ lies on the extended side of the mobility edge, while the state with $n = N_A + 1$ lies on the localized side. In Fig.~\ref{fig:IPR_Scaling_edge}(a), we plot $\ln(\mathrm{IPR})$ of the state $n=N_A$ as a function of $\ln(N)$ and extract a scaling exponent $D = 1.0 \pm 0.1$, confirming that this state is indeed extended. In Fig.~\ref{fig:IPR_Scaling_edge}(b), we plot $\ln(\mathrm{IPR})$ of the state $n=N_A+1$ as a function of $\ln(N)$ and extract a scaling exponent $D = 0.000 \pm 0.003$, indicating that this state is indeed localized. Thus we see that the transition between extended and localized states at the mobility edge is abrupt, with no intermediate phase. 
This behaviour should be contrasted with the results of the AA model, where the wave functions at the critical point are neither fully localized nor extended, but exhibit irregular multifractal fluctuations across all length scales \cite{Multi}.

\subsection{Structure of the eigenstates}

\begin{figure}[t]
  \centering
    \figlab{a}
  \includegraphics[width=0.55\textwidth]{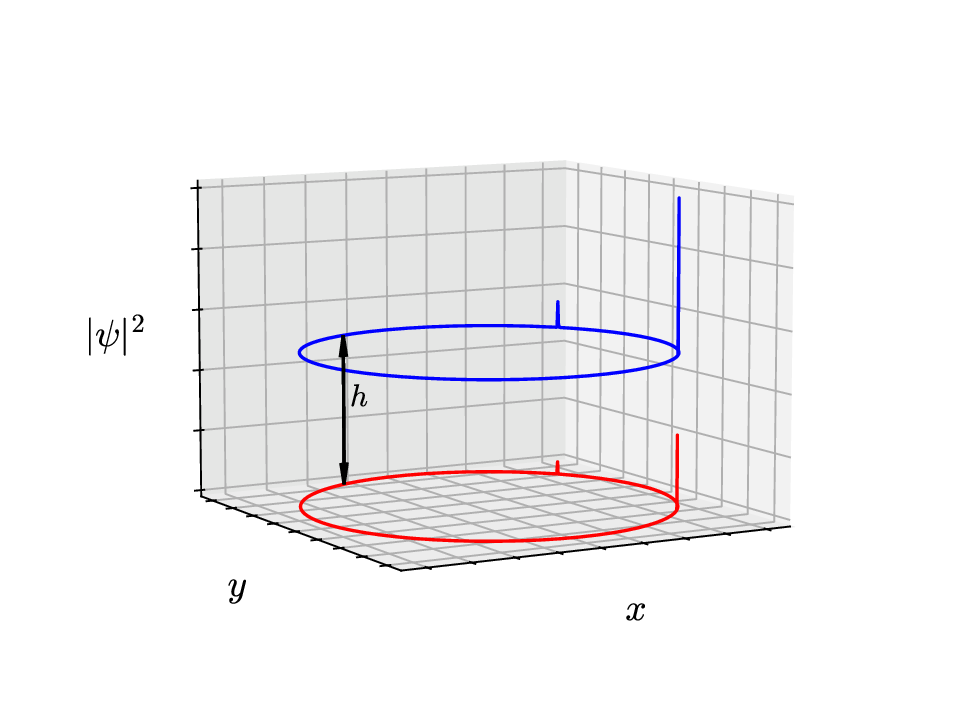}\\[1em]
    \figlab{b}
  \includegraphics[width=0.55\textwidth]{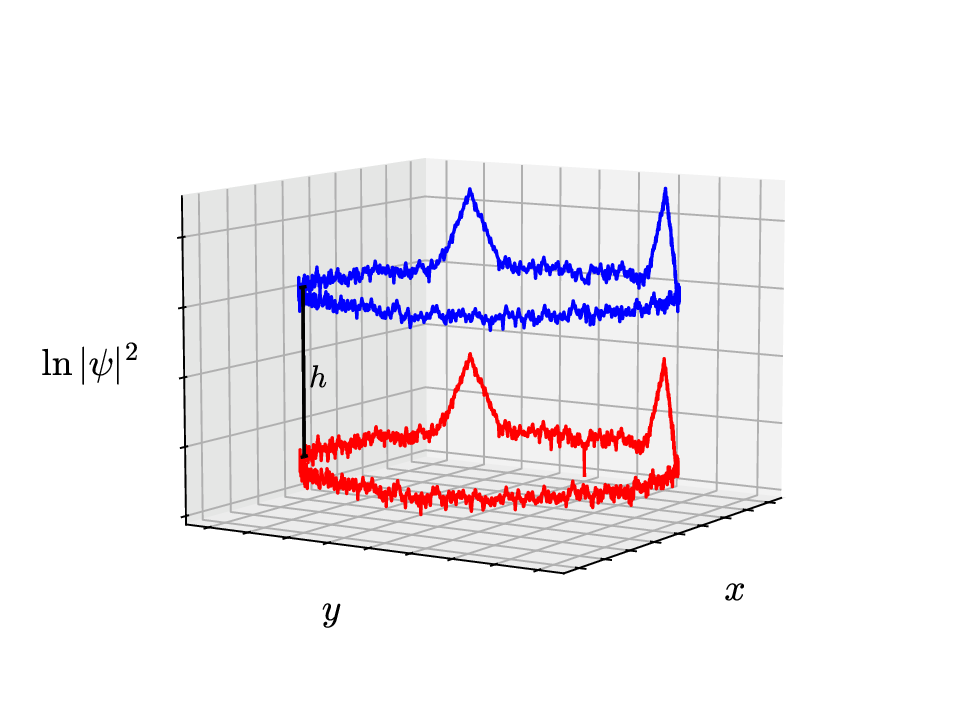}
  \caption{(Color online)
    (a) ``Bird's-eye'' view of the probability densities $|\psi_n(A,i_A)|^2$ and $|\psi_n(B,i_B)|^2$ in the localized state with $n=2750$, for $a=2$ and $d=1$. The smaller peaks have been scaled up in order to make them visible. (b) Logarithmic plot of the same probability densities without scaling.
    }
  \label{fig:combined}
\end{figure}

To illustrate the shape of the localized states, in Fig. \ref{fig:combined} we plot the probability densities $|\psi_n(A,i_A)|^2$ and $|\psi_n(B,i_B)|^2$ using as an example the state with $n=2750$. The wave function exhibits sharply localized peaks in both chains in close proximity of each other. To characterize its spatial extent of the main peak, we approximate the probability density in each chain by an exponential function of the form
\begin{equation}
    |\psi(\theta)|^2=|\psi(\theta_0)|^2\exp\left({-\frac{|\theta-\theta_0|}{\delta}}\right).
    \label{Eq: fit}
\end{equation}
Here, we used the angular coordinate $\theta$ to take advantage of the cylindrical symmetry of the system  and introduced  the angular localization length $\delta$. For convenience, we also define the linear localization length $\xi = R\delta$, allowing a direct comparison of the localization length with the lattice spacing. 

\begin{figure}[t]
\includegraphics[width=0.4\textwidth]{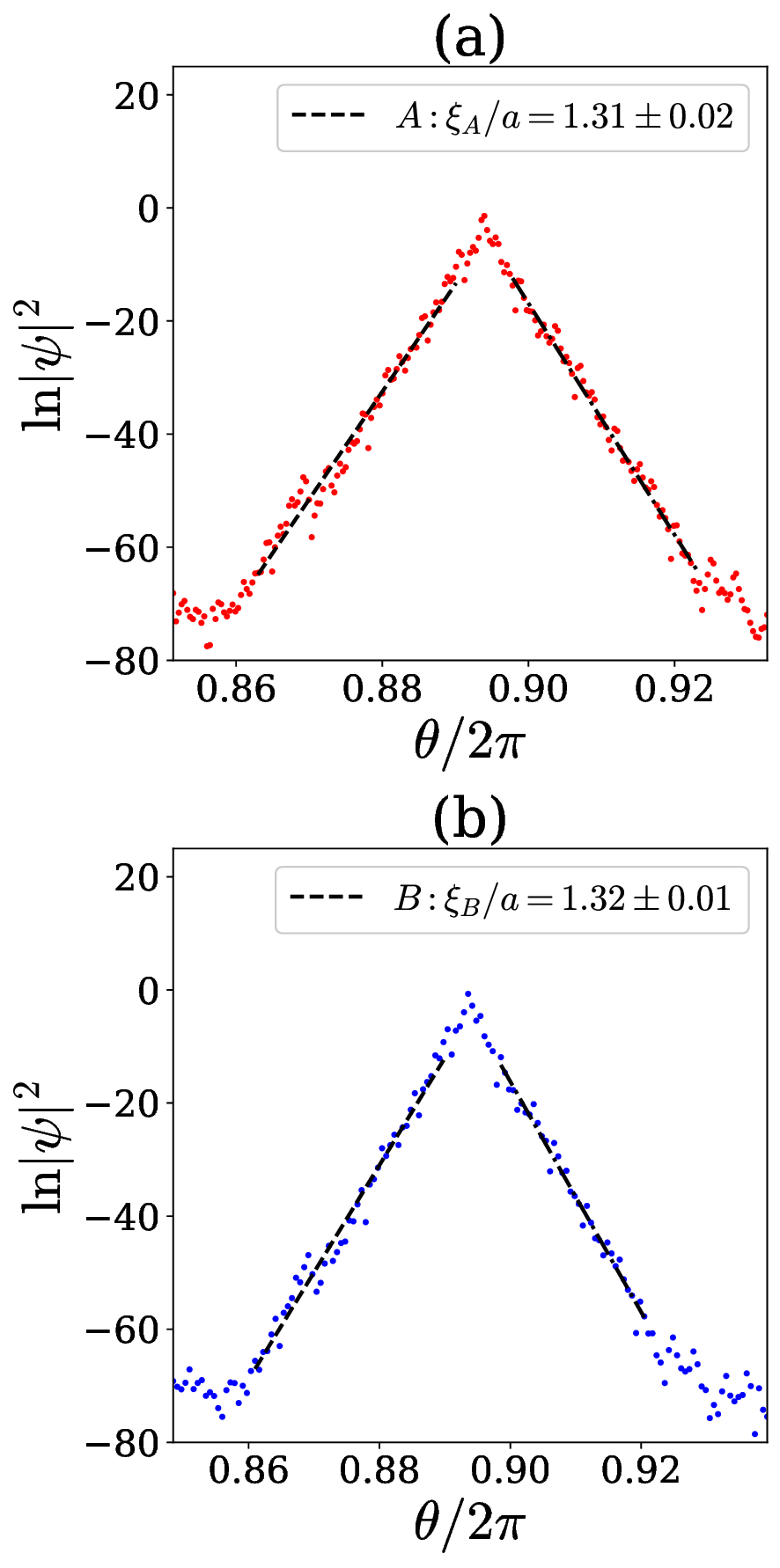}
  \caption{(Color online) Logarithm of the probability densities in the $A$ and $B$ chains for the $n=2750$ state ($a=2$ and $d=1$). The data is fitted to a straight line on each side of the peak. }
  \label{fig:log_psi}
\end{figure}

To explore the detailed structure of the localized eigenstates, in Fig.~\ref{fig:log_psi} we plot the logarithm of the probability density over the angular coordinate $\theta$ for the $n=2750$ state. The wave function has two peaks, which nearly mirror each other in both chains. The shorter peaks correspond to a much smaller probability density, on the order of $e^{-10}$. We extract the localization lengths in the $A$ and $B$ chains by averaging the linear fits on each side of the main peaks, as shown in Fig.~\ref{fig:log_psi}. For the $A$-chain, we find a localization length of $\xi_A/a = 1.31 \pm 0.02$, and for the $B$-chain we find $\xi_B/a = 1.32 \pm 0.01$. Thus, the localization lengths in the $A$ and $B$ chains are nearly identical.

\subsection{Effects of magnetic field}
\label{sec: results with B}

For a magnetic field $\bm{B} = B_\parallel \hat{\bm{z}}$, with the field strength characterized by the parameter~(\ref{Eq:chiz}), we observe no significant qualitative change in the shape of the IPR distribution for any choice of parameters. This indicates that localization is not affected by this orientation of the field.

\begin{figure}[t]
\includegraphics[width=0.48\textwidth]{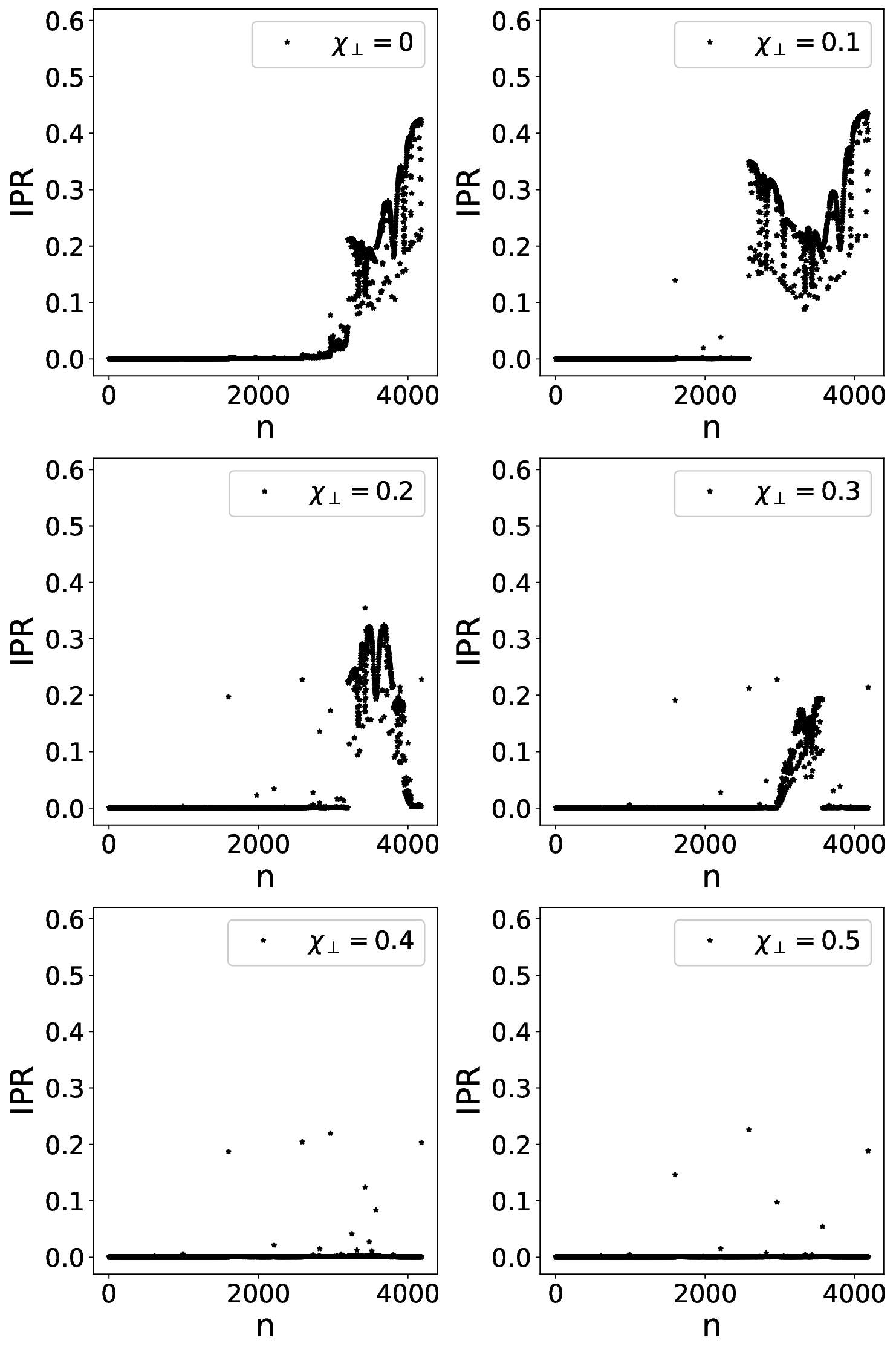}
  \caption{Values of the IPR as the strength of a perpendicular field $\chi_\perp$, as defined in Eq. (\ref{Eq:chir}), is increased for $a=1$ and $d=0.4$.}
  \label{fig:chir}
\end{figure}

In contrast, a magnetic field perpendicular to the plane of the chains, which corresponds to $\bm{B}=B_\perp\hat{\bm{r}}$ in the cylindrical geometry, has a pronounced effect on localization. In Fig. \ref{fig:chir} we show the evolution of the IPR as the magnetic field strength, characterized by the parameter~(\ref{Eq:chir}), increases. Initially, for small fields, the IPR values for some states increase, indicating a strengthening of localization in already localized states and the localization of some previously delocalized sites. However, as the field strength increases further, the majority of IPR values decrease to zero. This suggests that while weak fields can enhance localization in certain states, stronger fields lead to the delocalization of most states. Eventually, all states become delocalized.


\section{Conclusions}
\label{sec: conclusion}

In conclusion, we showed that electron states can be localized in a quasi-1D system without disorder, due to inter-chain hopping between two coupled incommensurate chains. In contrast to other models of incommensurate 1D crystals, such as the AA model, we observe the mobility edges separating extended from localized states, which are controlled by the inter-chain separation. These mobility edges are abrupt and coincide with gaps in the energy spectrum. Localization arises when inter-chain hopping dominates intra-chain hopping; however, this dependence is complex and not solely determined by the ratio of the distance between the chains to the lattice constant.

Previous work \cite{similar1} has studied mobility edges in 1D models with explicitly imposed incommensurate modulation of the off-diagonal hopping. In that model, the localized states are exponentially localized and the localization transition is accompanied by an intermediate critical phase. In contrast, in our model the incommensurate hopping modulation arises naturally from the geometry of two coupled periodic lattices, without additional site-dependent parameters. With this different physical mechanism, we do not observe an intermediate critical phase.

Our model is well suited for studying the effect of a magnetic field on localization, which we found to strongly depend on the field's orientation. A magnetic field parallel to the chains' plane does not alter localization properties. In contrast, a magnetic field perpendicular to the chains' plane enhances localization at small field strengths, at the same time localizing some previously extended states. At higher field strengths, most states become delocalized.

Experimentally, polariton setups have been used to emulate tight-binding models \cite{Polariton1, Polariton2}, including realizations of the AA model as demonstrated in Ref. \cite{Polariton3}. These setups utilize a planar semiconductor micro-cavity placed between two Bragg reflectors, where coupling between photons and electron–hole pairs creates exciton–polaritons. The latter represent the particles in a tight-binding model (these particles are neutral bosons). The micro-cavity is patterned into an array of identical etched micropillars, each supporting a single localized polariton mode and thus playing the role of the tight-binding lattice sites. Exponential decay of the hopping amplitudes arises naturally when the energy of the polariton mode is lower than the barrier created by the etched regions, with the decay length determined by the etching depth, which makes it possible to control hopping strength. 
These systems can be used to check predictions of our model, as they allow direct measurement of the wave function and can be extended to probe multiple coupled tight-binding chains with exponentially decaying hopping. Furthermore, one can synthesize an effective gauge potential in the polariton setups~\cite{Lim2017,Aidelsburger2018,Widmann2025}, making it possible to test our predictions about the effects of a magnetic field on the localization in incommensurate chains.

Our work admits several natural theoretical extensions. For instance, one can study the impact of hopping anisotropy due to $p$-wave or $d$-wave orbitals on the localization properties, as well as the differences between exponential and Gaussian hoppings. In order to better understand the transport properties of the system, one can investigate time evolution of wave packets. It would also be interesting to explore the many-body localization \cite{many-body} in our system. However, this would tremendously increase the Hilbert space dimension, which is not computationally feasible with our exact diagonalization method.


\acknowledgments

The authors are grateful to C. Wilson for useful discussions.
This work was supported by a Discovery Grant 2021-03705 from the Natural Sciences and Engineering Research Council of Canada.


\appendix

\section{Weakly coupled chains}
\label{app: decoupled chains}

In this appendix, all length scales are dimensionless, i.e. measured in the units of the wave function size $\lambda$. The wave vectors are also dimensionless, measured in the units of $\lambda^{-1}$. External magnetic field is equal to zero.

\begin{figure}[t]
\includegraphics[width=0.4\textwidth]{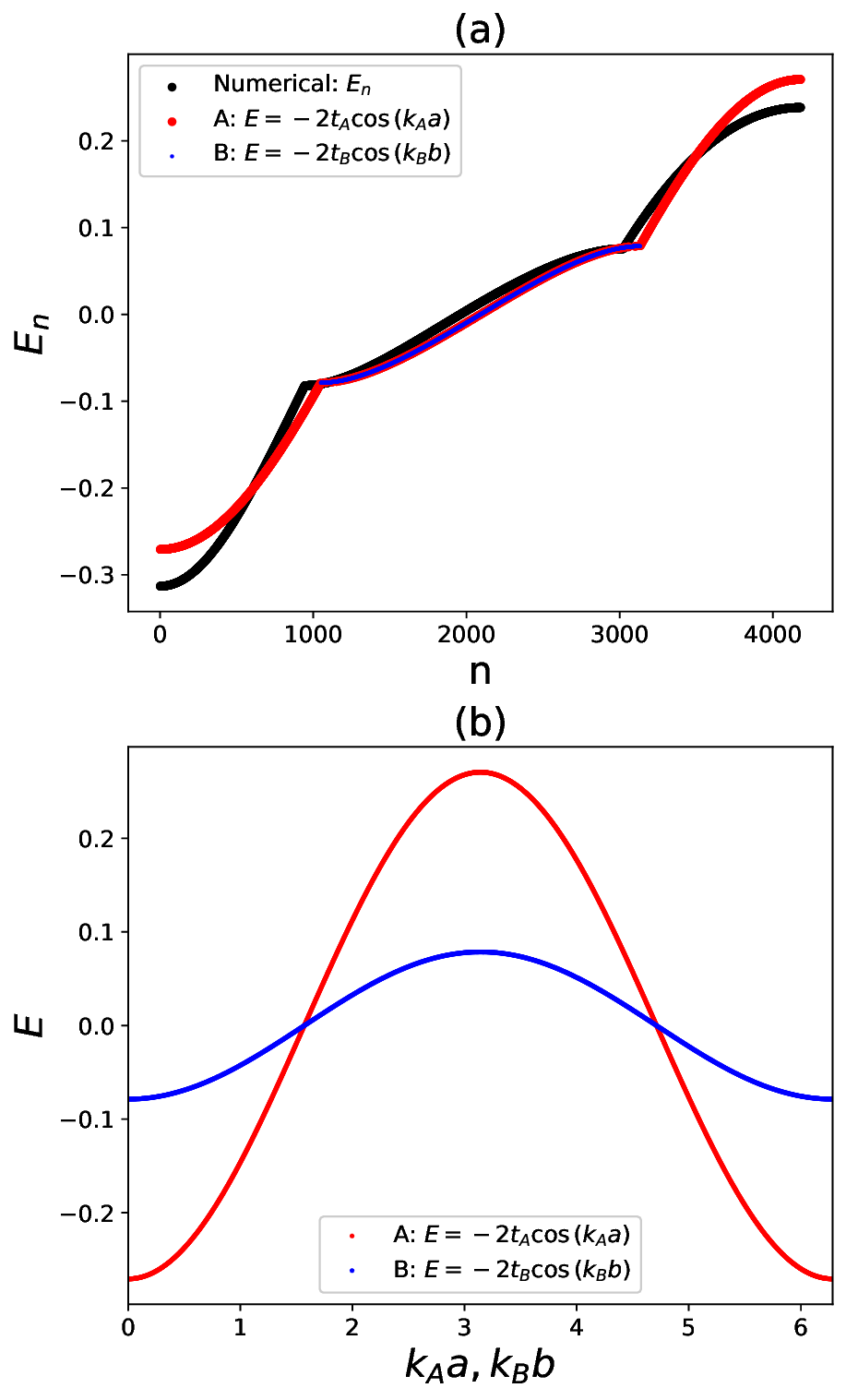}
  \caption{(Color online) Energy spectra for decoupled chains. (a) Numerical diagonalization of the Hamiltonian (\ref{eq:H-def}) for $a=2$ and $d=10$. (b) Analytical tight-binding spectrum, Eq. (\ref{eq:E-kAkB}), with $t_A=0.135$ and $t_B=0.039$.}
  \label{fig:decoupled}
\end{figure}

Let us set the inter-chain separation $d$ to a sufficiently large value, effectively decoupling the chains. As expected, we observe a smooth energy spectrum without gaps, and all values of the IPR become effectively zero. This result is presented in Fig.~\ref{fig:decoupled}(a).

At $d\to\infty$, the energy spectrum can be calculated analytically. The Hamiltonian (\ref{eq:H-def}) takes the form $\hat{H}=\hat H_0=\hat{H}_A+\hat{H}_B$, so that there are two independent families of plane-wave eigenstates:
\begin{eqnarray}
    && \ket{A,k_A}=\frac{1}{\sqrt{N_A}}
    \left(e^{ik_Aa},\dots,e^{ik_AN_Aa},0,\dots,0\right)^\top, \nonumber \\
    && \ket{B,k_B}=\frac{1}{\sqrt{N_B}}
    \left(0,\dots,0,e^{ik_Bb},\dots,e^{ik_BN_Bb}\right)^\top,\qquad
\label{Eq:eigenstates-decoupled}
\end{eqnarray}
where 
\begin{eqnarray}
\label{eq: k_Ak_B-quantized}
    && k_A=\frac{2\pi m_A}{N_Aa},\quad 0 \leq m_A \leq N_A-1, \nonumber\\
    && k_B=\frac{2\pi m_B}{N_Bb},\quad 0 \leq m_B \leq N_B-1,
\end{eqnarray}
and $N_Aa=N_Bb=L\gg 1$ is the dimensionless circumference of the chains. In the limit $N_A,N_B\to\infty$,
the wave vectors $k_A$ and $k_B$ corresponding to physically non-equivalent states become quasi-continuous variables limited to the ``first Brillouin zones'' (BZ): $0\leq k_A<2\pi/a$ and $0\leq k_B<2\pi/b$.

Although our numerical analysis retains all hopping terms, the ratios of next-nearest to nearest-neighbor hopping are typically quite small (for $a=2$, they are approximately equal to $0.135$ and $0.039$ in the $A$ and $B$ chains, respectively), so using the nearest-neighbor approximation in the  analytical calculation is well justified. Then, the dispersion relations corresponding to the eigenstates (\ref{Eq:eigenstates-decoupled}) take the form
\begin{equation}
\label{eq:E-kAkB}
    E^{(0)}_{A,k_A} = -2t_A\cos\bigl(k_A a\bigr),\
    E^{(0)}_{B,k_B} = -2t_B\cos\bigl(k_B b\bigr),
\end{equation}
which are plotted in Fig.~\ref{fig:decoupled}(b). When these energy values are ordered from lowest to highest, they closely match the numerical spectrum in Fig.~\ref{fig:decoupled}(a). The two ``tail''  regions on the left and right sides of Fig.~\ref{fig:decoupled}(a) contain only energies corresponding to the $A$ chain while the center region contains to points corresponding to both the $A$ and $B$ chains.   

As $d$ is decreased, we treat the effects of the inter-chain coupling perturbatively \cite{Sakurai2011}. The diagonal matrix elements of $\hat{\tilde H}$ in the basis (\ref{Eq:eigenstates-decoupled}) are equal to zero, which means that there are no first-order (or any odd order) corrections to the energy levels due to the inter-chain coupling. 
For even-order corrections we need to know the off-diagonal matrix elements. Using Eqs. (\ref{eq:t-dimensionless}) and (\ref{eq: k_Ak_B-quantized}) we obtain:
\begin{eqnarray}
     && \bra{A,k_A}\hat{\tilde H}\ket{B,k_B} = \bra{B,k_B}\hat{\tilde H}\ket{A,k_A}^*\nonumber\\ 
     && \quad = \frac{1}{\sqrt{N_AN_B}}\sum_{i_A=1}^{N_A}\sum_{j_B=1}^{N_B}e^{-ik_Ai_Aa}e^{ik_Bj_Bb}\,\tilde t(i_A,j_B)\ \nonumber\\
     && \quad = \frac{1}{\sqrt{N_AN_B}}\sum_{\theta_A,\theta_B}e^{-im_A\theta_A}e^{im_B\theta_B} F(\theta_A-\theta_B),\qquad 
     \label{Eq:off_diag}
\end{eqnarray}
where $\theta_A=2\pi i_A/N_A$ and $\theta_B=2\pi j_B/N_B$ are the angular positions of atoms in the chains, and 
$$
    F(\Theta)=\exp\Biggl[-\sqrt{\left(\frac{L}{\pi}\right)^2\sin^2\frac{\Theta}{2}+d^2}\Biggr]
$$
is a $2\pi$-periodic function. Representing the latter as a Fourier series, $F(\Theta)=\sum_M F_M e^{iM\Theta}$, and using the identity
$$
    \sum_{j=1}^N \exp\left[\frac{2\pi i(m-M)}{N}j\right]=N\delta_{m,M\,\mathrm{mod}\,N},
$$
we calculate the $\theta$ sums in Eq. (\ref{Eq:off_diag}) and obtain:
\begin{eqnarray*}
    && \bra{A,k_A}\hat{\tilde H}\ket{B,k_B} \\
    && \quad = \sqrt{N_AN_B}\sum_M F_M \delta_{m_A,M\,\mathrm{mod}\,N_A} \delta_{m_B,M\,\mathrm{mod}\,N_B}.
\end{eqnarray*}
Therefore, the matrix elements (\ref{Eq:off_diag}) are nonzero only if the wave vectors satisfy the condition
\begin{equation}
\label{eq:k_A-k_B-condition}
    k_A+\frac{2\pi}{a}n_A=Q=k_B+\frac{2\pi}{b}n_B,
\end{equation}
where $n_A$ and $n_B$ are integers, and $Q=2\pi M/L$.
If $n_A=n_B=0$, then the transitions between the plane-wave states (\ref{Eq:eigenstates-decoupled}) caused by the inter-chain hopping conserve momentum, i.e. $k_B=k_A$. If either $n_A$ or $n_B$ is nonzero, then the transitions correspond to ``Umklapp'' processes, in which $k_B$ is not equal to $k_A$. 

If the incommensurability ratio $\rho$ is irrational, e.g., equal to the golden ratio $\phi$, then Eq. (\ref{eq:k_A-k_B-condition}) has an infinite number of solutions for $k_B$ at any given $k_A$. To show this, let us introduce $\nu_A=k_Aa/2\pi$ and $\nu_B=k_Bb/2\pi$, which satisfy $0\leq\nu_{A,B}<1$. Then, we obtain from Eq. (\ref{eq:k_A-k_B-condition}):
\begin{equation}
\label{eq:XY}
    Y=\rho X,
\end{equation}
where $X=n_A+\nu_A$ and $Y=n_B+\nu_B$. Fixing $\nu_A$ and going through all integer values of $n_A$, one obtains an infinite set of solutions $Y_{n_A}(\nu_A)$ of the above equation. Their fractional parts $\{Y_{n_A}(\nu_A)\}$ give the values of $\nu_B$, while their integer parts correspond to $n_B$. For finite chains, $\rho$ is equal to $N_A/N_B$---a ratio of two large co-prime numbers. Then, the number of solutions for $\nu_B$ of Eq. (\ref{eq:XY}) is finite and equal to $N_B$.   

\begin{figure}[t]
\includegraphics[width=0.4\textwidth]{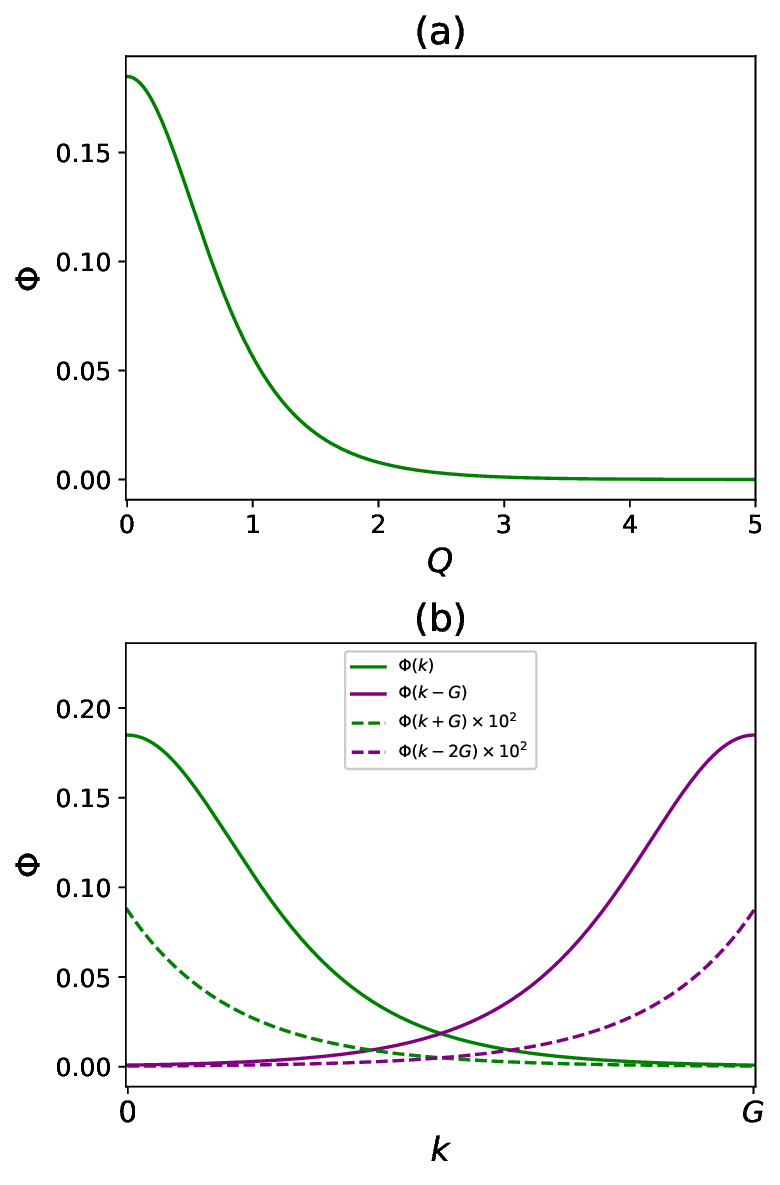}
\caption{(Color online) Magnitudes of the inter-chain matrix elements, Eq. (\ref{eq:off-diag-final}), for $d=1.5$ and $a=2$. (a) $\Phi(Q)$ decays rapidly as $Q$ increases. (b) $\Phi(k+Gn)$ plotted in the BZ, $0 \leq k < G$. Note the difference in scale between the largest ($n=0,-1$) and next-largest ($n=1,-2$) matrix elements. 
}
\label{fig:Phi}
\end{figure}

Although the number of nonzero matrix elements at given $k_A$ is large or even infinite, their magnitudes, which depend on $M$ and therefore on $Q$, are expected to decrease rapidly as the Umklapp parameters $n_A$ and $n_B$ increase. These magnitudes are determined by the Fourier coefficients $F_M$, which are evaluated as follows:
\begin{eqnarray*}
    F_M &=& \int_{-\pi}^\pi\frac{d\Theta}{2\pi}\,e^{-iM\Theta} F(\Theta) \\
    &\simeq& \int_{-\infty}^\infty\frac{d\Theta}{2\pi}\,e^{-iM\Theta} \exp\Biggl[-\sqrt{\left(\frac{L}{2\pi}\right)^2\Theta^2+d^2}\Biggr] \\
    &=& \frac{2d}{L\sqrt{1+Q^2}}K_1\left(d\sqrt{1+Q^2}\right).
\end{eqnarray*}
In the second line here we used the fact that at $L\gg 1$ the function $F(\Theta)$ is sharply peaked near $\Theta=0$, which allows one to replace $\sin(\Theta/2)\to\Theta/2$ and extend the integration limits to infinities. Then the integral can be calculated analytically \cite{GR-6ed}, with $K_1(x)$ being the modified Bessel function of the second kind.
Therefore, we obtain for the matrix elements (\ref{Eq:off_diag}):
\begin{eqnarray}
\label{eq:off-diag-final}
    \braket{A,k_A|\hat{\tilde H}|B,k_B} \simeq \frac{2d^2}{\sqrt{ab}}\sum_Q \Phi(Q) \qquad \nonumber \\
    \qquad \times \delta_{Q,k_A+2\pi n_A/a} \delta_{Q,k_B+2\pi n_B/b},
\end{eqnarray}
where the function
$$
    \Phi(Q) = \frac{1}{d\sqrt{1+Q^2}}K_1\left(d\sqrt{1+Q^2}\right) 
$$
is plotted in Fig. \ref{fig:Phi}(a).

Assuming that the separation between the chains is sufficiently large to make the perturbative treatment of the inter-chain hopping quantitatively valid, one can use the large-$x$ asymptotics $K_1(x)\simeq\sqrt{\pi/2x}\,e^{-x}$ \cite{AS-book}.
The exponential decay of $\Phi(Q)$ means that the matrix elements rapidly decrease as $|Q|$ increases. According to Eq. (\ref{eq:off-diag-final}), this constrains the possible values of the Umklapp numbers $n_A$ and $n_B$. Namely, by far the largest matrix elements correspond to $n_A,n_B=0$ or $-1$, and the next largest (and considerably smaller) -- to $n_A,n_B=1$ or $-2$. This is illustrated in Fig. \ref{fig:Phi}(b), where the magnitudes $\Phi(k+Gn)$ are plotted as functions of momentum $k$ in the BZ $0\leq k<G$, for different values of $n$ (here $k=k_A$ or $k_B$, $G=2\pi/a$ or $2\pi/b$, and $n=n_A$ or $n_B$).

Substituting $n_A,n_B=0$ or $-1$ in the constraint (\ref{eq:k_A-k_B-condition}), and using the fact that $k_A$ and $k_B$ are restricted to their respective BZs, it is easy to check that there are only two possibilities:
\begin{equation}
\label{eq:k_A-k_B-1}
    k_A=k_B,
\end{equation}
which corresponds to $n_A=n_B=0$, and
\begin{equation}
\label{eq:k_A-k_B-2}
    k_A-\frac{2\pi}{a}=k_B-\frac{2\pi}{b},
\end{equation}
which corresponds to $n_A=n_B=-1$. For $k_A$ and $k_B$ not satisfying these conditions, the matrix elements (\ref{eq:off-diag-final}) either vanish identically or have a negligibly small magnitude for $d\gtrsim 1$. 

We can now calculate the corrections to the $A$-chain dispersion. The second-order correction takes the form
\begin{equation}
\label{Eq:secon_order}
    E^{(2)}_{A,k_A}=\sum_{k_B } \frac{|\braket{A,k_A| \hat{\tilde H} |B,k_B}|^2}{E^{(0)}_{A,k_A}- E^{(0)}_{B,k_B} },
\end{equation}
where the possible values of $k_B$ are determined by Eqs.~(\ref{eq:k_A-k_B-1}) and (\ref{eq:k_A-k_B-2}). The above expression diverges when the condition
\begin{equation}
\label{eq:inter-band-degeneracy}
     E^{(0)}_{A,k_A} =E^{(0)}_{B, {k_B} }
\end{equation}
is satisfied. The locations of these inter-band degeneracies depend on the parameters of the system. For example, for $a=2$ the solutions of Eq. (\ref{eq:inter-band-degeneracy}) with $k_A$ and $k_B$ related through Eq. (\ref{eq:k_A-k_B-1}) or (\ref{eq:k_A-k_B-2}) are given by $k_A^{\text{gap}}a \simeq 1.526$, $1.861$, $4.422$, and $4.757$.
The divergence of $E^{(2)}_{A,k_A}$ at these points indicates that we must turn to the degenerate perturbation theory, in which the degeneracies are removed by splitting the energy bands near the intersection points. These spectral gaps are not visible in Figs. \ref{fig:4gstop} and \ref{fig:En_evolution} where the energy levels are arranged from lowest to highest.

The spectral gaps shown in Figs. \ref{fig:4gstop} and \ref{fig:En_evolution} originate from the \textit{intra-band} degeneracies. To see this, we bias one chain, say the $B$ chain, by adding a constant on‑site potential term to $\hat H_B$. For a sufficiently strong bias, the dispersion relations in the $A$ and $B$ chains no longer overlap and the condition (\ref{eq:inter-band-degeneracy}) is never satisfied, yet the gaps persist in the energy spectrum according to our numerical results. 
To capture this effect, we must include the next nonzero term in the perturbation theory, namely the fourth‑order energy correction. Using the procedure for deriving the higher-order perturbative corrections developed in Refs. \cite{Huby_1961,Silver1970} we obtain:  
\begin{widetext}
\begin{eqnarray}
    E^{(4)}_{A,k_A} = \sum_{k_A'\neq k_A}\sum_{k_B,k_B'} \frac{ \braket{A,k_A|\hat{\tilde H}|B,k_B} \braket{B,k_B|\hat{\tilde H}|A,k'_A} \braket{A,k'_A|\hat{\tilde H}|B,k'_B} \braket{B,k'_B|\hat{\tilde H}|A,k_A} }{(E^{(0)}_{A,k_A}-E^{(0)}_{B,k_B}) (E^{(0)}_{A,k_A}-E^{(0)}_{A, k'_A }) (E^{(0)}_{A,k_A}-E^{(0)}_{B,k'_B})} \nonumber\\
   - E^{(2)}_{A,k_A}  \sum_{k_B}\frac{|\braket{A,k_A| \hat{\tilde H} |B,k_B}|^2}{(E^{(0)}_{A,k_A}- E^{(0)}_{B,k_B})^2 }.
   \label{Eq:fourth_order}
\end{eqnarray}
\end{widetext} 
In addition to the already detected inter-chain degeneracies described by Eq. (\ref{eq:inter-band-degeneracy}), which are removed by biasing the $B$ chain, the denominator in the first term vanishes when the condition
\begin{equation}
    E^{(0)}_{A,k_A} =E^{(0)}_{A,k'_A }
\label{eq:E_A_E_A-prime}
\end{equation}
is satisfied for $k_A'\neq k_A$. 
In Eq. (\ref{Eq:fourth_order}), the values of $k_B$, $k'_A$, and $k'_B$ at given $k_A$ allowed by the conditions (\ref{eq:k_A-k_B-1}) and (\ref{eq:k_A-k_B-2}) are given by
\begin{eqnarray}
\label{eq:k_BAB}
    && k_B=k_A,\ k_A-\frac{2\pi}{a}+\frac{2\pi}{b}, \nonumber\\
    && k_A'=k_A,\ k_A-\frac{2\pi}{a}+\frac{2\pi}{b},\ k_A+\frac{2\pi}{a}-\frac{2\pi}{b}, \qquad \\
    && k_B'=k_A,\ k_A-\frac{2\pi}{a}+\frac{2\pi}{b}. \nonumber
\end{eqnarray}
It is easy to check that for other values of $k_B'$ one cannot complete the closed-loop virtual hopping sequence
$$
    \ket{A,k_A} \rightarrow \ket{B,k_B} \rightarrow \ket{A,k'_A} \rightarrow \ket{B,k'_B} \rightarrow \ket{A,k_A},
$$
as enforced by the chain of matrix elements in the first term of Eq.~(\ref{Eq:fourth_order}). 

According to Eqs.~(\ref{eq:E_A_E_A-prime}), (\ref{eq:E-kAkB}), and (\ref{eq:k_BAB}), gaps in the spectrum may open up at the wave vectors satisfying the following equation:
\begin{equation*}
    \cos(k_A a) = \cos\!\left(k_A a + 2\pi \mathfrak{n}_1  + \frac{2\pi a}{b}\mathfrak{n}_2\right),
\end{equation*}
where $(\mathfrak{n}_1,\mathfrak{n}_2)=(1,-1)$ or $(-1,1)$. The solutions have the form $k^{\text{gap}}_A = -(\pi/b)\mathfrak{n}_2 + (\pi/a)\mathfrak{n}$, where $\mathfrak{n}$ is an integer. 
Finally, using Eq. (\ref{rho-def}) we obtain for the energy gap locations in the BZ:
\begin{equation}
\label{eq:k_A-GAP}
    \frac{k_A^{\text{gap}}a}{2\pi}=\frac{1}{2}-\frac{1}{2\rho},\; \frac{1}{2\rho},\; 1-\frac{1}{2\rho},\; \frac{1}{2}+\frac{1}{2\rho}.
\end{equation}
Thus there are two pairs of gaps located symmetrically about the center of the spectrum, the ``outer'' ones corresponding to the first and fourth values of $k_A^{\text{gap}}$ above, and the ``inner'' ones corresponding to the second and third values of $k_A^{\text{gap}}$.  
It should be noted that this method does not provide the energy gap width, but merely indicates where in momentum space a gap may occur due to the failure of non-degenerate perturbation theory in the fourth order.

\begin{figure}[t]
\includegraphics[width=0.4\textwidth]{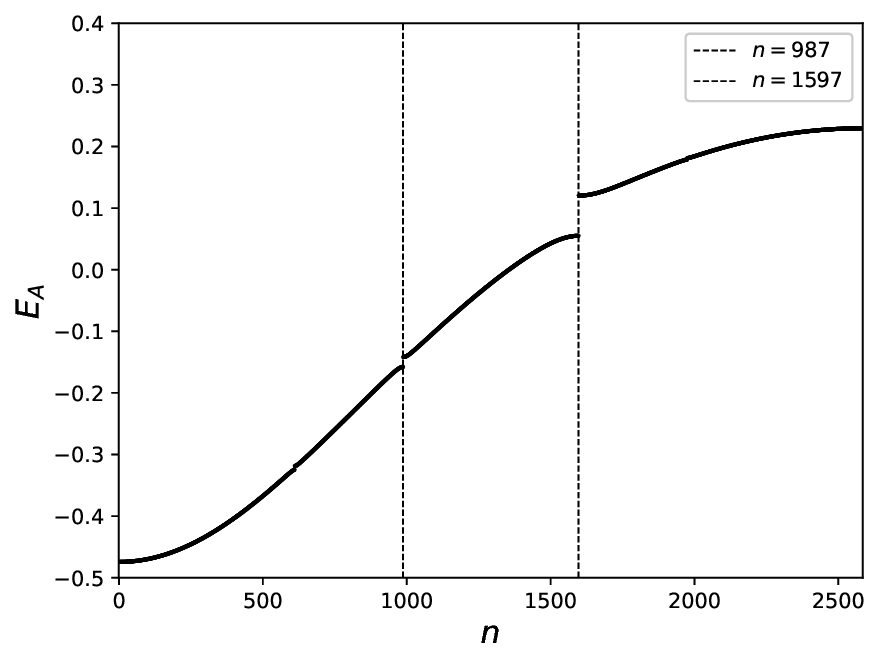}
\caption{Numerical spectrum of Eq. (\ref{Eq:Eigen}) for $a=2$ and $d=1$, showing only the $A$-chain band modified by coupling to the $B$ chain. The part corresponding to the modified $B$-chain band is shifted to higher energies by a strong uniform potential and is not shown. The black dashed lines mark the gap locations predicted by Eq. (\ref{eq:E_A_E_A-prime}).}
\label{fig:GAPS}
\end{figure}

\begin{figure}[t]
  \centering
  \figlab{a}\\[0.5ex]
  \includegraphics[width=0.4\textwidth]{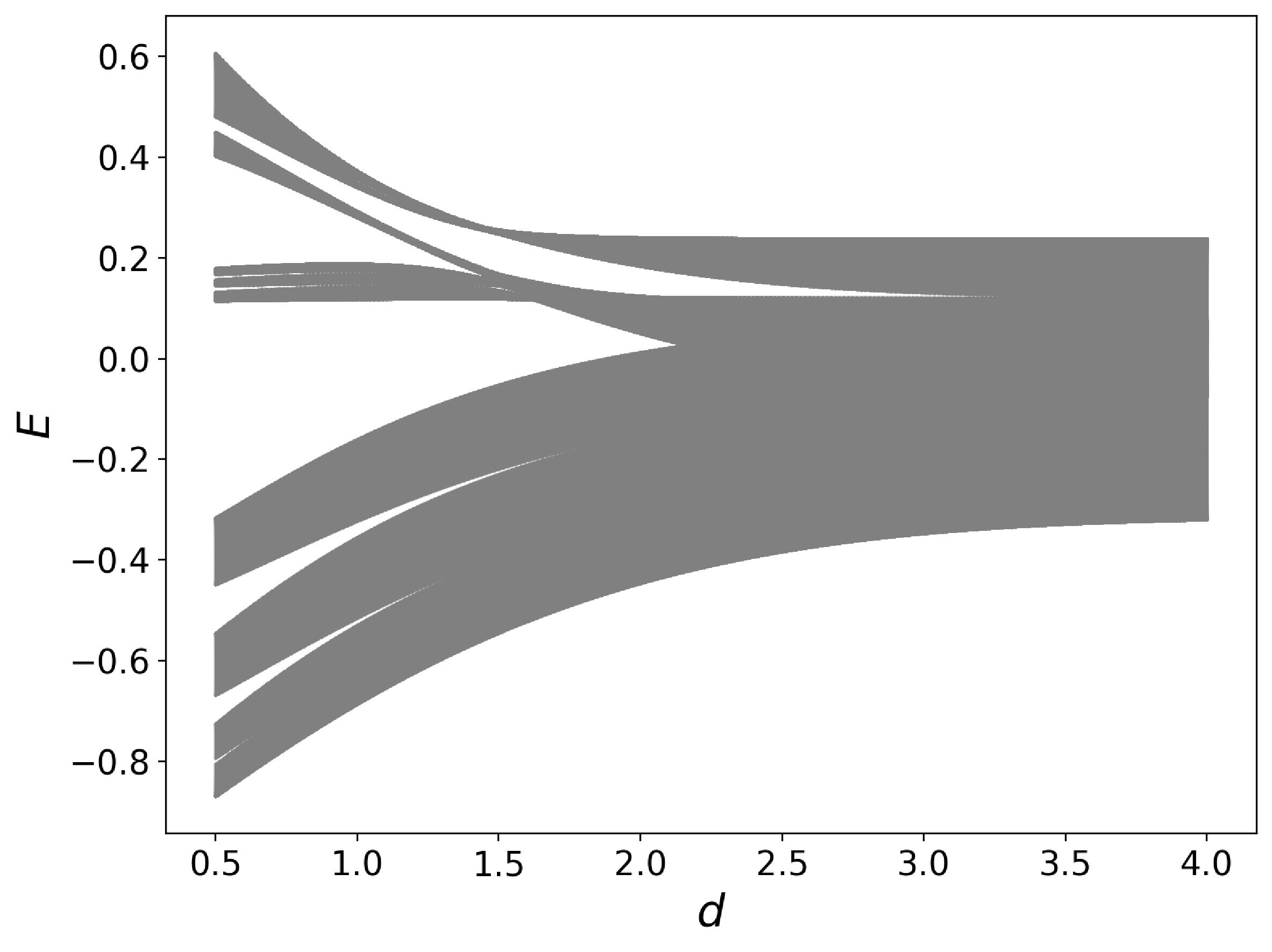}\\[1em]
  \figlab{b}\\[0.5ex]
  \includegraphics[width=0.4\textwidth]{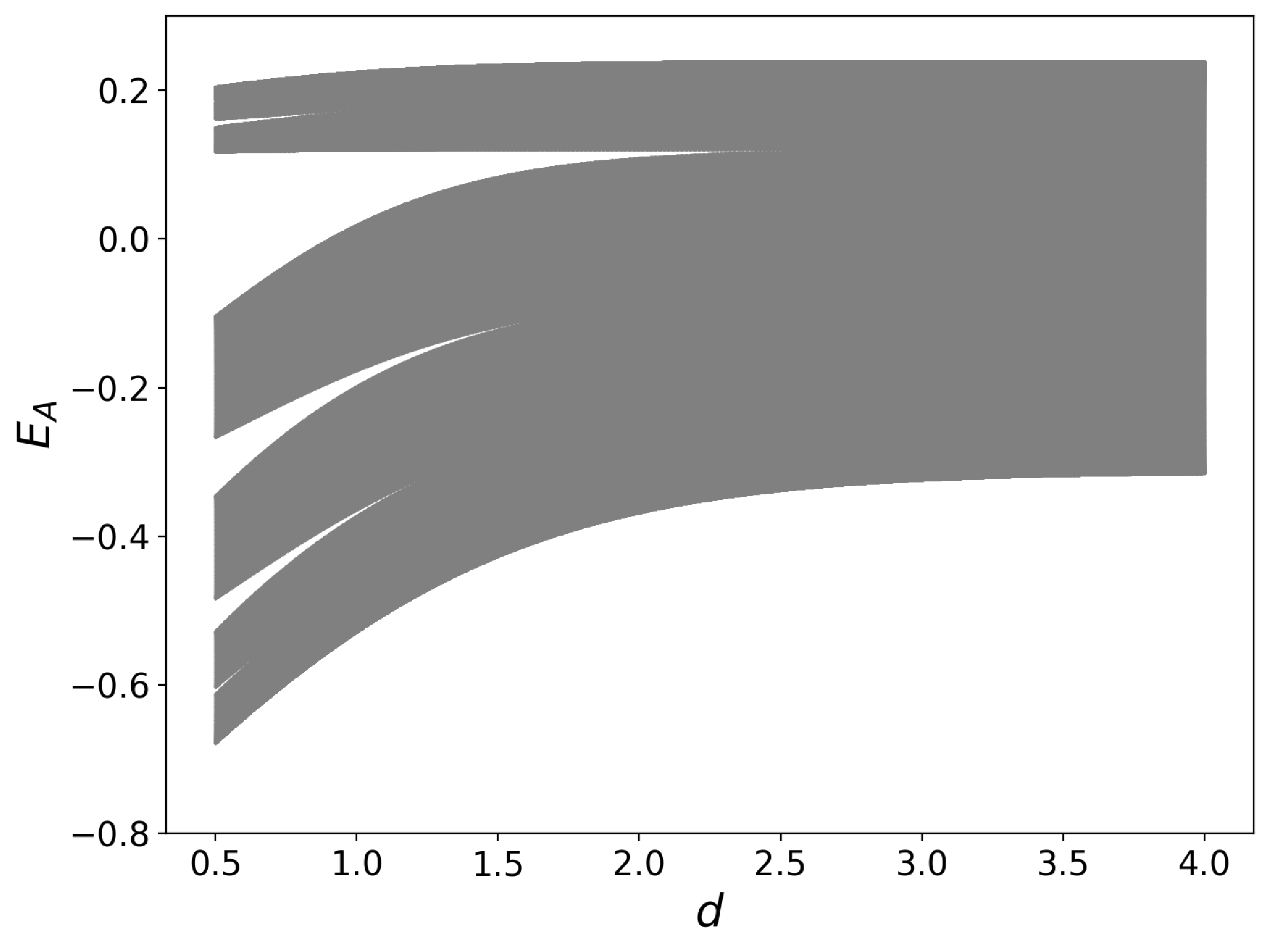}
  \caption{(a) Numerical spectrum of the model~(\ref{eq:H-def}), with $N_A=2584$, $N_B=1597$, and $a=1$, as $d$ is varied.
(b) Numerical spectrum of the same model with the $B$ chain biased by a uniform potential, showing only the $A$-chain band modified by coupling to the $B$ chain. }
  \label{fig:E_d_plots}
\end{figure}

In our numerical analysis in Sec. \ref{sec: results}, the eigenvalues are ordered from lowest to highest according to an integer index $n$. To compare the gap structure predicted by perturbation theory applied to the spectrum (\ref{eq:E-kAkB}) with these numerical results, we arrange the perturbed spectrum in the same manner. The gaps in momentum space given by (\ref{eq:k_A-GAP}) appear symmetrically about the center of the spectrum, which means that each state in the ordered spectrum actually corresponds to two states in the original spectrum. It follows from Eqs. (\ref{eq: k_Ak_B-quantized}) and (\ref{rho-def}) that the number of states with energies below the outer gaps is equal to $N_A-N_B$, while the number of states with energies below the the inner gaps is equal to $N_B$.
These numbers give the locations of the gaps in the ordered spectrum at indices
\begin{equation}
\label{eq:n-gaps-A-band}
    n = N_A-N_B,\ N_B.
\end{equation}
Substituting here our values for $N_A$ and $N_B$, we find gaps located at $n = 987$ and $1597$. This prediction agrees exactly with our numerical results in the case where the $B$ chain is biased such that $E^{(0)}_{B,k_B} \gg E^{(0)}_{A,k_A}$. The lower part of the spectrum corresponding to the $A$-chain dispersion modified by the coupling to the $B$ chain is displayed in Fig. \ref{fig:GAPS}. The upper part of the spectrum, which originates from the biased $B$ chain, is pushed up to higher energies and is not shown. 
To illustrate the evolution of spectral gaps as the inter-chain distance is varied, in Fig.~\ref{fig:E_d_plots} we plot the numerical energy spectrum as a function of $d$.

\bibliography{literature}

\end{document}